\documentclass[%
reprint,
superscriptaddress,
nofootinbib,
amsmath,amssymb,
aps,
pra,
noeprint,
longbibliography
]{revtex4-2}

\usepackage{graphicx}
\usepackage[english]{babel}
\usepackage{appendix}
\usepackage{dcolumn}
\usepackage{bm}
\usepackage{color}
\usepackage{booktabs,eqparbox}
\usepackage{braket}
\usepackage{dsfont}
\usepackage{placeins}
\usepackage{paralist}
\usepackage{comment}
\usepackage[utf8]{inputenc}
\usepackage{pgfplots}
\usepackage{amsmath}
\usepackage{graphicx}
\usepackage{lipsum}
\usepackage{tikz}
\usepackage{csquotes}
\usepackage{hyperref}
\usepackage{cleveref}
\usepackage{textgreek}
\usepackage[printonlyused,withpage]{acronym}
\usepackage{quantikz}
\usepackage{ifthen}

\DeclareUnicodeCharacter{2212}{−}
\usepgfplotslibrary{groupplots,dateplot}
\usetikzlibrary{patterns,shapes.arrows}
\pgfplotsset{compat=newest}

\makeatletter
\let\oldtheequation\theequation
\renewcommand\tagform@[1]{\maketag@@@{\ignorespaces#1\unskip\@@italiccorr}}
\renewcommand\theequation{(\oldtheequation)}
\makeatother

\definecolor{myRed}{HTML}{A3061E}
\definecolor{myBlue} {RGB} {0,63,119}
\definecolor{myYellow} {cmy} {0,0.263,0.741}
\definecolor{myGreen}{HTML}{0B6E4F}
\definecolor{FAU}{RGB}{0,56,101}
\definecolor{gray75}{gray}{0.75}
\colorlet{myOrange} {myYellow!60!myRed}
\colorlet{myViolet}{myRed!50!myBlue!80}

\renewcommand{\vec}[1]{\bm{#1}}

\newcommand{\mycomment}[1]{}

\RequirePackage[normalem]{ulem} 
\RequirePackage{color}\definecolor{RED}{rgb}{1,0,0}\definecolor{BLUE}{rgb}{0,0,1} 

\addto\extrasenglish{%
}

\addto\extrasenglish{%
}

\begin{document}

\title{Variational Time Evolution Compression for Solving Impurity Models on Quantum Hardware}

\author{Stefan Wolf}
\affiliation{Department of Physics, Friedrich-Alexander Universität Erlangen-Nürnberg, Erlangen, Germany}
\author{Martin Eckstein}
\affiliation{Institute of Theoretical Physics, University of Hamburg, 20355 Hamburg, Germany}
\affiliation{The Hamburg Centre for Ultrafast Imaging, Luruper Chaussee 149, 22761 Hamburg, Germany}

\author{Michael J. Hartmann}
\affiliation{Department of Physics, Friedrich-Alexander Universität Erlangen-Nürnberg, Erlangen, Germany}

\date{\today} 

    \begin{abstract}
        Dynamical mean-field theory (DMFT) is a useful tool to analyze models of strongly correlated fermions like the Hubbard model. In DMFT, the lattice of the model is replaced by a single impurity site embedded in an effective bath. The resulting single impurity Anderson model (SIAM) can then be solved self-consistently with a quantum-classical hybrid algorithm. This procedure involves repeatedly preparing the ground state on a quantum computer and evolving it in time to measure the Green's function. We here develop an approximation of the time evolution operator for this setting by training a Hamiltonian variational ansatz. The parameters of the ansatz are obtained via a variational quantum algorithm that utilizes a small number of time steps, given by the Suzuki-Trotter expansion of the time evolution operator, to guide the evolution of the parameters. The resulting circuit has a fixed depth for the time evolution depending on the size of the bath and is significantly shallower than a comparable Suzuki-Trotter expansion. 
    \end{abstract}

\maketitle

\section{Introduction}
The simulation of quantum many-body systems poses a formidable challenge in diverse scientific disciplines, from condensed matter physics to quantum chemistry and material science.
Classical computation methods struggle to efficiently model the complex entanglement and correlations inherent in such systems.
This caused the development of methods to treat these systems in simplified form, such as dynamical mean-field theory (DMFT) \cite{Georges} for strongly correlated fermions, which is based on a local approximation for the many-body self-energy. Within DMFT,
observables of a lattice model are obtained from a simpler quantum impurity model, which consists of only one or few interacting orbitals embedded in a continuous effective bath of non-interacting fermions. 
However, in contrast to a conventional mean-field theory, the underlying impurity model of DMFT remains a many-body system, as the dynamics of the bath is still taken into account. 

To find a solution of the quantum impurity model using a Hamiltonian based solver, the bath is typically discretized into a finite number of bath sites, which are each interacting with the impurity site. The quality of the results depends on the number of bath sites such that the solution of the impurity model is again suffering from an exponential growth of the Hilbert space. 
The success of DMFT has largely profited from the availability of efficient quantum Monte Carlo (QMC) algorithms for impurity models \cite{Gull}. However, QMC is mostly formulated in imaginary time and, therefore, does not give access to spectral information or even real-time dynamics \cite{Aoki2014}. 
This issue becomes even more rampant in the extension of DMFT to Cluster-DMFT, where a cluster of sites is used instead of a single impurity site \cite{Lichtenstein, Kotliar, Potthoff2003, Degenfeld} where QMC algorithms can face a sign problem even in equilibrium.
The advantage of Cluster-DMFT is that it also gives access to spatial correlation between sites in the cluster impurity, whereas DMFT for single-site impurities is not able to account for such effects by construction. 
This difference becomes important in the low-dimensional case of real materials and application for which single-site DMFT is at most a good approximation. 
Cluster DMFT can therefore be viewed as an intermediate step between DMFT and the simulation of the original many-body problem, while also increasing the computational costs drastically.
It is important to note that state-of-the-art classical impurity solvers based on the numerical renormalization group (NRG) and density matrix renormalization group (DMRG) have become very powerful real-frequency or real-time impurity models, which have been successfully extended to multiorbital calculations at low or zero temperature \cite{Wolf, Bauernfeind2017, Stadler} and realistic material simulations \cite{Bauernfeind2018, Kugler}.
But these solvers also suffer from the exponential growth of the Hilbert space with the number of bath orbitals. 
When simulating real time dynamics, the entanglement of the system grows linearly with time, causing the required computational resources to grow exponentially as well.

Quantum computing offers a promising paradigm for overcoming these restrictions.
DMFT is naturally suitable for a hybrid approach where the impurity model is solved using the quantum device for finding the Green's function of the impurity site or, in a more matured form, the impurity cluster. 
The solution is then fed back into a classical calculation to determine the self-consistent bath, and its representation in terms of a finite number of bath orbitals. The DMFT loop between classical and quantum computer is performed until convergence is reached. 

The task performed by the quantum device to obtain the Green's function can vary but is in general in one of two categories. 
The first category consists of algorithms that rely on measuring observable in the ground state \cite{Eckstein}. 
The measurements are then used to reconstruct the Green's function, mostly using its Lehmann representation \cite{Endo, Rungger, Jamet, Rizzo, Ehrlich, Selisko}. 
The algorithms of the second category rely on measuring the Green's function for different points in time and therefore require performing time evolution on the quantum hardware \cite{Bauer, Kreula, Jaderberg, Keen, Steckmann}.
The latter can be accomplished by using a Trotterization of the time evolution operator \cite{Suzuki, Yoshida, Barthel, Ostmeyer}, or other approximations 
\cite{Low2017, Low2019, Gilyen, Cirstoiu, Martyn, Gibbs, Zhao, Mansuroglu2, Mansuroglu}. 

Both categories require that the ground state is prepared on the quantum device. 
There are different methods to achieve this, including quantum phase estimation \cite{kitaev}, algorithms incorporating imaginary time evolution \cite{McArdle} or cooling methods by utilizing ancilla qubits as a fridge to remove energy from the system \cite{Lin, Puente, Marti}. 
The current common choice, however, is to prepare the ground state by using a variational quantum eigensolver (VQE) \cite{Cerezo2014, Peruzzo, Tilly}. 

In the current early stage of quantum computing, most of the effort related to DMFT is focused on developing proof of concept algorithms for single-site impurities.
Previous work mostly concentrated on using the quantum computer for solving a simplified version of DMFT with just a single bath site (called two-site DMFT) at half-filling \cite{Endo, Rungger, Jamet, Rizzo, Ehrlich, Bauer, Kreula, Jaderberg, Keen, Steckmann} and recent work showed results for up to three bath sites on quantum hardware \cite{Hogan}.
In this work, we focus on evaluating the Green's function using the time evolution and extend previous work to the general case, away from half-filling and with multiple bath sites. 
We present an improved version of the algorithm for compressing the time evolution into a shallow quantum circuit introduced in \cite{Jaderberg, Benedetti, Barison, Berthusen, Puig, Goh} to make it suitable for application in generic cases. 
We further show that there exists an upper limit on the required circuit depth that depends on the number of bath sites for the ground state preparation and for the approximation of the time evolution operator for the considered system sizes. 
To further reduce the required time evolution, we implement a post processing strategy that allows a reconstruction of the Green's function via the Lehmann representation.
We investigate the performance of our algorithm for systems that are also solvable on classical hardware. While we do not claim to surpass state-of-the-art classical methods, our objective is to establish a framework for addressing larger systems, such as cluster DMFT, on near-term quantum devices.

\section{Dynamical Mean Field Theory on a Quantum Device}
A simple model to describe strongly correlated fermions on a lattice is the Hubbard model \cite{Hubbard},
\begin{align}
\begin{split}
    \hat{H}_\text{Hub} &= U \sum_{i} \hat{n}_{i,\uparrow}\hat{n}_{i,\downarrow} - \mu \sum_{i,\sigma} \hat{n}_{i,\sigma} \\
    &+ \sum_{\sigma} \sum_{\braket{i,j}} v_{ij} (\hat{c}^\dag_{i,\sigma}\hat{c}_{j,\sigma} + \text{h. c.} )
    \label{HubbardHamiltonian}
\end{split}
\end{align}
where $U$ is the on-site interaction strength, $\mu$ the chemical potential and $v_{i,j}$ the hopping amplitude between the lattice sites $i$ and $j$, which we choose to be equal for all pairs. 
$\hat{c}_{i,\sigma}^\dag$ $\hat{c}_{i,\sigma}$ are fermionic creation and annihilation operators and $\hat{n}_{i,\sigma} = \hat{c}_{i,\sigma}^\dag\hat{c}_{i,\sigma}$ is the number operator, acting on the spin orbital $\{i,\sigma\}$. Here, we only consider nearest neighbor hopping (expressed by $\braket{i,j}$ in the summation) with the same spin. 
Despite its simplicity, an analytic solution of this model is only known for the one dimensional case \cite{Essler}. For higher dimensions, numerical simulations of the systems quickly reach the computational limits of current classical computing, due to  the exponential increase of the Hilbert space dimension. 

Dynamical mean-field theory (DMFT) is a method to analyze the Hubbard model with less computational resources. 
Here, the Hubbard Hamiltonian is mapped onto an impurity model by singling out one lattice site and replacing all other sites by an effective bath. 
The impurity site can then exchange fermions with the bath such that the problem remains a many-body problem in contrast to a classical mean field theory. 
The mapping of the lattice problem onto the impurity model is justified by the observation that the lattice self-energy becomes local and therefore does no longer depend on momentum $\mathbf{k}$,
\begin{equation}
    \Sigma_\text{latt}(\mathbf{k},\omega) \overset{d \rightarrow \infty}{\rightarrow} \Sigma_\text{latt}(\omega),
    \label{selfenergyconsistency}
\end{equation}
in the limit of high dimensions, $d$;
$\Sigma_\text{latt}(\omega)$ is then obtained from the self-energy $\Sigma_\text{imp}(\omega)$ in the impurity model.
The central quantities of DMFT are then the single-particle retarded Green's function of the lattice and the impurity,
\begin{align}
    G_\text{latt}^\text{R}(\omega) &= \int\limits_{-\infty}^{+\infty}d\epsilon\frac{\rho(\epsilon)}{\omega+\mu-\Sigma_\text{latt}(\omega)-\epsilon}
    \label{localGreensFunction}\\
    G_\text{imp}^\text{R}(\omega) &= \frac{1}{\omega + \mu - \Delta (\omega) - \Sigma_\text{imp}(\omega)} \label{impurityGreensFunction},
\end{align}
where $\rho(\epsilon)$ is the non-interacting density of states of the lattice model and $\Delta(\omega)$ is called hybridization, which describes the exchange of fermions between the impurity and the bath. 
The problem at hand is solved if the self consistency condition, 
\begin{equation}
    G_\text{latt}^\text{R}(\omega) = G_\text{imp}^\text{R}(\omega),
    \label{greensFunctionconsistency}
\end{equation}
is fulfilled.

The first step of DMFT involves mapping the lattice problem onto a suitable impurity model with a bath discretized into an infinite number of bath orbitals, each with different on-site energy.
In practice, of course, the number of bath orbitals is limited by the computational resources.
Different Hamiltonian representations of the impurity model can be chosen as long as they allow fermion exchange between the bath and the impurity site.  A common choice is the ``star-shaped'' single impurity Anderson model (SIAM), the Hamiltonian of which consists of three terms,
\begin{equation}
    \hat{H}_\text{SIAM} = \hat{H}_\text{imp} + \hat{H}_\text{hyb} + \hat{H}_\text{bath},
    \label{SIAMHamiltonian}
\end{equation}
with
\begin{align}
    \hat{H}_\text{imp} &= U\hat{n}_{d,\uparrow}\hat{n}_{d,\downarrow} - \mu(\hat{n}_{d,\uparrow}+\hat{n}_{d,\downarrow}),\\
    \hat{H}_\text{hyb} &= \sum_{p=1}^B \sum_{\sigma \in\{\uparrow,\downarrow\}}V_{p}(\hat{d}^\dag_{\sigma}\hat{c}_{p,\sigma} + \text{H. c.}),\\
    \hat{H}_\text{bath} &=\sum_{p = 1}^B\sum_{\sigma\in\{\uparrow,\downarrow\}} \epsilon_p \hat{c}^\dag_{p,\sigma}\hat{c}_{p,\sigma}.
\end{align}
 $\hat{H}_\text{imp}$ describes the dynamics of the impurity site for both spins, while $\hat{H}_\text{bath}$ consists of the terms describing the bath sites and the exchange of fermions between both subsystems is given by $\hat{H}_\text{hyb}$. Here, $\hat{d}_{\sigma}$ and $\hat{d}^\dag_{\sigma}$ are the fermionic creation and annihilation operators of the impurity site with spin $\sigma$, $\hat{c}_{p,\sigma}^\dag$ and $\hat{c}_{p,\sigma}$ are the corresponding operators for the bath sites and $\hat{n}_{d,\sigma}=\hat{d}_{\sigma}^\dag\hat{d}_{\sigma}$ is the number operator.
The interaction between the two spin sites of the impurity, $U$, and the chemical potential, $\mu$, are predetermined by the underlying Hubbard model, but the other parameters, $V_p$ and $\epsilon_p$ must be adjusted until condition \ref{selfenergyconsistency} is fulfilled. 
As one can see in the Hamiltonian of the SIAM, the problem remains a many-body problem in DMFT.

The self-consistent solution is then conveniently  formulated via the noninteractiung Green's function of the impurity model
\begin{equation}
    \Lambda(\omega) = \left(G^{R}(\omega)^{-1} + \Sigma(\omega)\right)^{-1},
    \label{Dysonequation}
\end{equation}
which is also called ``Weiss function'' in some analogy to the Weiss field in conventional mean-field theory. For the discrete bath model given by Eq.~\eqref{SIAMHamiltonian}, we have
\begin{equation}
    \Lambda(\omega) = \omega + \mu -\Delta(\omega), \label{WeissFunction}
\end{equation}
where the hybridization $\Delta(\omega)$ is a function of $V_p$ and $\epsilon_p$ only,
\begin{equation}
    \Delta(\omega) = \sum_p \frac{V_p^2}{\omega - \epsilon_p}
    \label{hybridization}
\end{equation}
Equation \ref{Dysonequation} allows the following  recursive algorithm to determining the self-energy self-consistently:
\begin{enumerate}\setcounter{enumi}{0}
    \item Make a initial guess of $\Sigma_0(\omega)$
    \item Calculate $G_\text{latt}^\text{R}(\omega)$ with equation \ref{localGreensFunction}
    \item Use equation  \ref{Dysonequation} to calculate the Weiss function $\Lambda(\omega)$
    \item Through $\Lambda(\omega)$ the parameters of the impurity model (\ref{SIAMHamiltonian}) are obtained via equation  \ref{WeissFunction}
    \item Solve the impurity model (\ref{SIAMHamiltonian}) to get $G_\text{imp}^\text{R}(\omega)$
    \item Calculate the new self-energy $\Sigma_n(\omega)$ using equation  \ref{Dysonequation} again and repeat steps 2 to 6 until $|\Sigma_{n-1}(\omega)-\Sigma_{n}(\omega)| < \delta$ for some threshold $\delta$
\end{enumerate}
We can further simplify the mapping procedure further by considering the underlying lattice of the Hubbard model to be a Bethe lattice with infinite connectivity, corresponding to semi-elliptic density of states $\rho(\epsilon)=\sqrt{4 v^2 - \epsilon^2}/(2\pi v^2)$ with bandwidth $4v$ ($v=1$ will be used as an energy unit below). For the Bethe lattice, the hybridization function can be expressed directly in terms of the Green's function
\begin{equation}
    \Delta(\omega) = v^2G_\text{imp}(\omega).
    \label{bethemapping}
\end{equation}
In this case, the DMFT loop reduces to the quantum simulation of $G_{imp}$ (step 5), and the determination of the bath parameters from $\Delta(\omega)$ (step 4).

In step (4), the determination of the new set of SIAM parameters  from the hybridization function $\Delta(\omega)$ is in practice not done in the real, but rather in the imaginary frequency domain, by introducing a fictive finite temperature $T = \frac{1}{\beta}$ and the Matsubara frequencies, 
\begin{equation}
    \omega_n = \frac{(2n+1)\pi}{\beta}.
\end{equation}
The Matsubara Green's function can then be calculated by the transformation,
\begin{equation}
    G_\text{imp}(i\omega_n) = \int_{-\infty}^\infty d\omega \frac{A_\text{imp}(\omega)}{i\omega_n - \omega},
\end{equation}
where we have introduced the spectral function,
\begin{equation}
    A_\text{imp}(\omega) = -\frac{1}{\pi}\text{Im}G_\text{imp}^\text{R}(\omega+i\eta)
\end{equation}
for an infinitesimal positive $\eta$.
In practice, the number of bath sites is always limited to a small  number $B$. 
Therefore, the hybridization is approximated by a truncated version, i.e. $\Delta(\omega)^B \approx \Delta(\omega)$ with the number of bath sites $B$. This approximation motivates the use of the cost function
\begin{equation}
    d = \frac{1}{n_\text{max}}\sum_{n=0}^{n_\text{max}} \left| v^2G_\text{imp}(i\omega_n) - \Delta(i\omega_n)^B\right|^2
    \label{mappingcost}
\end{equation}
which, once minimized, returns the best possible set of SIAM parameters for the next iteration. 
Nevertheless, the restriction in the number of bath sites for classical calculations will worsen the quality of the approximation. 
Performing step (5) in the DMFT loop with the help of a quantum device holds the promise to lift this restriction, by allowing for an increase in the number of bath orbitals (and impurity orbitals in the case of cluster DMFT), provided the quantum device is large and powerful enough. 

An important quantity, which we will use to verify the accuracy of our results, is the quasiparticle weight $\mathcal{Z}$, 
\begin{equation}
    \mathcal{Z} = \frac{1}{1 - \left.\frac{d\text{Re}(\Sigma(\omega+i\eta))}{d \omega}\right|_{\omega=0}} \, ,
\end{equation}
which is the spectral weight of the quasiparticle peak.
Since the evaluation of the quasiparticle peak is numerical unstable as it depends on the derivative of the difference of two inverse quantities (see Eq. \ref{Dysonequation}), we will instead estimate $\mathcal{Z}$ using the Matsubara self energy,
\begin{equation}
    \mathcal{Z}_\text{mats} = \frac{1}{1-\left.\frac{d\text{Im}(\Sigma(i\omega_n)}{d\omega_n}\right|_{\omega_n\rightarrow 0}}.
\end{equation}
Since this quantity is evaluated with a finite temperature, it introduces an error compared to the zero-temperature quantity $\mathcal{Z}$.
This error increases with $U$ \cite{Georges}, but it gives a sufficient indication of the quality of the result since the mapping to the SIAM is also done in the imaginary frequency space to increase numerical stability.

\section{Method}
\subsection{Jordan Wigner Transformation}
Solving the problem at hand involves the calculation of the Matsubara Green's function $G_\text{imp}^\text{R}(\omega)$. One way to do so is calculating the retarded Green's function  $G_\text{imp}^\text{R}(t)$, followed by a Fourier transformation. 
Here, we use a hybrid quantum-classical algorithm to evaluate the time depended Green's function
\begin{equation}
    G^\text{R}_{\text{imp}, \sigma}(t) = -i\Theta(t)(\braket{\hat{c}_{0,\sigma}(t)\hat{c}^\dag_{0,\sigma}(0)} + \braket{\hat{c}^\dag_{0,\sigma}(0)\hat{c}_{0,\sigma}(t)})
    \label{timedependentGreensFunction}
\end{equation}
The Fourier transformation is then performed classically, where, for zero-temperature, the thermal expectation values reduce to expectation values with respect to the ground state. 
For $T>0$ these expectation values must be evaluated with respect to a Gibbs state. While it is in principle possible to prepare such a state using e. g. thermofield double states \cite{Zhu} or a variational quantum thermaliser \cite{Selisko_Thermalizer}, this adds additional computational resources. 
Even though our algorithm may be extended for this more interesting use case, we only consider the zero-temperature limit in this work.

\begin{figure}
\centering
\begin{quantikz}
    \lstick{$\ket{0}_\text{anc}$} & \gate{\hat{H}} & \ctrl{1} & \qw  & \ctrl{1} & \gate{\hat{H}} & \meter{} \\
    \lstick{$\ket{GS}$} & \qw & \gate{\hat{X}_{0,\sigma}} & \gate{\hat{U}(t)} & \gate{\hat{X}/\hat{Y}_{0,\sigma}} & \qw & \qw 
\end{quantikz}
\caption{Measuring the Green's function w.r.t. the ground state on a quantum device. Using the Hadamard test and measuring $\braket{Z_{\text{anc}}}$ returns the desired expectation values.}
\label{GFmeasure}
\end{figure}
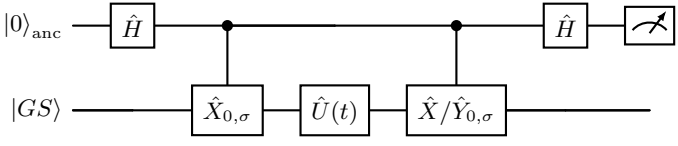

Performing this calculation on quantum hardware requires mapping the fermionic creation and annihilation operators onto operations which can be performed on the quantum computer. 
The different mappings are distinguished by their locality. 
Local mappings have a low Pauli weight but introduce additional qubits required for storing the parity \cite{Verstraete, Setia, Whitfield, Derby, Chen, OBrien}.
In contrast, nonlocal mappings require only as many qubits as fermionic modes, but a single operator can involve every qubit in the worst case \cite{Bravyi2002, Bravyi2017, Jiang, Vlasov, Miller}.
The most intuitive (and for this work the most beneficial) mapping is the Jordan-Wigner transformation (JWT) \cite{JordanWigner}.
We apply the transformation to each spin sector separately, which is possible, since every term in $\hat{H}_{\text{SIAM}}$, except for $\hat{H}_\text{imp}$, only acts on qubits representing the same spin.
The impurity sites are represented by the qubits with index $0$, while each bath site is represented by a qubit with the same index $p$ so that
\begin{align*}
    \hat{d}_\sigma^\dag &\mapsto \frac{1}{2}(\hat{X}_{0,\sigma}-i\hat{Y}_{0,\sigma})\\
    \hat{d}_\sigma &\mapsto \frac{1}{2}(\hat{X}_{0,\sigma}+i\hat{Y}_{0,\sigma})\\
    \hat{c}^\dag_{p,\sigma} &\mapsto \frac{1}{2}\hat{Z}_{0,\sigma}\hat{Z}_{1,\sigma}\dots\hat{Z}_{p-1,\sigma}\left(\hat{X}_{p,\sigma}-i\hat{Y}_{p,\sigma}\right),\\
    \hat{c}_{p,\sigma} &\mapsto \frac{1}{2}\hat{Z}_{0,\sigma}\hat{Z}_{1,\sigma}\dots\hat{Z}_{p-1,\sigma}\left(\hat{X}_{p,\sigma}+i\hat{Y}_{p,\sigma}\right),
\end{align*}
with the Pauli-Gates $\hat{X}_{i}$, $\hat{Y}_{i}$ and $\hat{Z}_{i}$ acting on qubit $i$. 
Depending on the ordering of the spins onto the qubits an additional string $\hat{Z}_{0,\sigma}\dots\hat{Z}_{p,\sigma}$ must be included in front of the operators for spin $\bar{\sigma}$.
Applying the JWT onto $\hat{H}_\text{SIAM}$ gives
\begin{align}
    \hat{H}_\text{imp} &=  \frac{U}{4}\hat{Z}_{0,\uparrow}\hat{Z}_{0,\downarrow}  + \left(\frac{\mu}{2} - \frac{U}{4}\right) \left(\hat{Z}_{0,\uparrow} + \hat{Z}_{0,\downarrow}\right),\\
    \hat{H}_\text{hyb} &= \sum_{p=1}^B\sum_{\sigma} \frac{V_{p,\sigma}}{2} \hat{X}_{0,\sigma}\hat{Z}_{1,\sigma}\dots\hat{Z}_{p-1,\sigma}\hat{X}_{p,\sigma}\\
    &+ \sum_{p=1}^B\sum_\sigma \frac{V_{p,\sigma}}{2}\hat{Y}_{0,\sigma}\hat{Z}_{1,\sigma}\dots\hat{Z}_{p-1,\sigma}\hat{Y}_{p,\sigma} \nonumber \\
    \hat{H}_\text{bath} &=-\sum_{p=1,\sigma}^B\frac{\epsilon_p}{2}\hat{Z}_{p,\sigma},
\end{align}
Due to the JWT, the Hamiltonian of the SIAM is now non-local and includes terms with actions on up to $B+1$ qubits. 
For implementation on quantum hardware, fermionic swap gates \cite{Verstraete2009} are useful to remove the large $\hat{Z}$-strings between the $\hat{X}/\hat{Y}$-pair acting on qubits $q_{0,\sigma}$ and $q_{p,\sigma}$. 
These switch the position of two qubits while conserving the parity. They do not increase the total number of two-qubit gates, since they can be incorporated into the hybridization terms (see appendix 
\ref{appendix:trotterization} for more details).
The JWT is also used to transform the impurity Green's function. 
After taking symmetry effects into account, Eq. \ref{timedependentGreensFunction} reduces to measuring two expectation values w.r.t. the ground state $\ket{GS}$ on the quantum device,
\begin{align}
\begin{split}
    G^\text{R}_{\text{imp},\sigma}(t) &= \frac{-i}{2} \left(\text{Re}(\braket{\hat{U}^\dag(t)\hat{X}_{0,\sigma}\hat{U}(t)\hat{X}_{0,\sigma}})\right.\\
    &\left.+i \text{Re}(\braket{\hat{U}^\dag(t)\hat{Y}_{0,\sigma}\hat{U}(t)\hat{X}_{0,\sigma}})\right)
    \label{timedependetGreensFunctionJWT}
\end{split}
\end{align}
Both expectation values can be measured using an ancilla qubit and a Hadamard test via the circuit shown in Fig.\ref{GFmeasure}.
The Hadamard test requires an ancilla qubit that is idling for most of the run time. 
This issue should be treated using e. g. dynamical decoupling to reduce the error on the ancilla qubit or a leakage gadget to determine if the ancilla qubit has leaked.
Reducing idling time is also advantageous. 
Our variational time evolution compression presented in section \ref{subsec:timeevolutioncompression} achieves this by requiring significantly less circuit depth to approximate time evolution compared to simple Trotterization (see section \ref{subsec:trotterization}).
Recent works propose alternative methods to measure the Green's function that require either also an idling qubit while applying the time evolution operator \cite{Bishop}, or additional Pauli operations during time evolution \cite{Piccinelli}.
Consequently, the latter approach results in a deep circuit that cannot be shortened by compression algorithms and is useful for quantum hardware beyond the NISQ era.  

\subsection{Trotterization}
\label{subsec:trotterization}
Evaluating Eq. \ref{timedependetGreensFunctionJWT} on a quantum device requires the application of the time evolution operator $\hat{U}(t) = e^{-it\hat{H}_\text{SIAM}}$.
To this end, a unitary generated by a Hamiltonian of the form
\begin{equation*}
    \hat{H} := \sum_{i=0}^M\hat{h}_{i}
\end{equation*}
can be approximated by a product of unitaries, each generated by one $\hat{h}_i$, in a Trotter Suzuki expansion.
This decomposition introduces an error, called Trotter error, if $[\hat{h}_i,\hat{h}_j]\neq 0$ for some $i,j$.
This error can be reduced by applying the unitaries repeatedly and alternating, i.e. for a large number $N$,
\begin{equation}
    \hat{U}(t) \approx \left(\prod_{i=0}^M e^{-i\frac{t}{N}\hat{h}_i}\right)^N =: \left(\hat{U}_\text{trotter}(\Delta t)\right)^N
    \label{trotterexpansion}
\end{equation}
with $\Delta t := \frac{t}{N}$. To reduce the Trotter error further, a higher order of the Trotter Suzuki expansion can be considered, e.g. for the second order:
\begin{equation}
    \hat{U}(t) \approx \left(\left(\prod_{i=0}^M e^{-i\frac{t}{2N}\hat{h}_{M-i}}\right)\left(\prod_{i=0}^M e^{-i\frac{t}{2N}\hat{h}_i}\right)\right)^N
    \label{higherordertrotterexpansion}
\end{equation}
In this case, the Trotter error scales with $\mathcal{O}\left(\frac{T^3}{N^2}\right)$. 
Here, we only consider the second order Trotterization of $\hat{H}_\text{SIAM}$. 
However, the terms of $\hat{H}_\text{hyb}$ require the application of additional Fermionic-swap gates. 
These gates can be incorporated into the gate structure at hand without increasing the gate count. 
Yet after just a single Trotter step, the lattice sites would no longer be represented by their respective qubit from the JWT. 
Especially the impurity sites would not be next to the ancilla qubit required for the Hadamard test.
We therefore pair two Trotter steps together and reveres the application order of the second hybridization term to restore the correct qubit representation. We therefore refer to 
\begin{equation}
    \hat{U}_\text{trotter}(\Delta t) := \left(e^{-\frac{i\Delta t}{4}\hat{H}_\text{imp+bath}}e^{-\frac{i\Delta t}{2}\hat{H}_\text{hyb}}e^{-\frac{i\Delta t}{4}\hat{H}_\text{imp+bath}}\right)^2
    \label{trotterstepdefinition}
\end{equation}
as a single Trotter Step in the remainder of this work (see appendix \ref{appendix:trotterization} for details).

\begin{figure*}
\centering
\begin{quantikz}
    \lstick[1]{$\ket{2,\sigma}$}\qw & \qw & \qw & \gate[wires = 2][1cm]{R_{XX+YY}} & \gate{R_Z} & \gate{R_{ZZ}} & \gate[wires = 2][1cm]{R_{XX+YY}} & \qw & \qw & \qw\\
    \lstick[1]{$\ket{1,\sigma}$}\qw & \qw & \gate[wires = 2][1cm]{R_{XX+YY}} & \qw & \gate{R_Z}& \qw & \qw & \gate[wires = 2][1cm]{R_{XX+YY}}& \qw & \qw\\
    \lstick[1]{$\ket{0,\sigma}$}\qw & \gate[wires = 2][1cm]{R_{ZZ}} & \qw & \qw & \gate{R_Z}& \qw & \qw & \qw &\gate[wires = 2][1cm]{R_{ZZ}} & \qw\\
    \lstick[1]{$\ket{0,\bar{\sigma}}$}\qw & \qw & \gate[wires = 2][1cm]{R_{XX+YY}} & \qw & \gate{R_Z}& \qw & \qw & \gate[wires = 2][1cm]{R_{XX+YY}}& \qw & \qw\\
    \lstick[1]{$\ket{1,\bar{\sigma}}$}\qw & \qw & \qw & \gate[wires = 2][1cm]{R_{XX+YY}} & \gate{R_Z}& \qw& \gate[wires = 2][1cm]{R_{XX+YY}} & \qw & \qw& \qw\\
    \lstick[1]{$\ket{2,\bar{\sigma}}$}\qw & \qw & \qw & \qw & \gate{R_Z} & \gate{R_{ZZ}} & \qw &\qw& \qw& \qw
\end{quantikz}
\caption{One layer of the second order Hamiltonian variational ansatz using the same two qubit gate count as $\hat{U}_\text{trotter}(\Delta t)$ for a system with two bath sites ($B=2$, see appendix \ref{appendix:trotterization} for details) }
\label{ansatz}
\end{figure*}
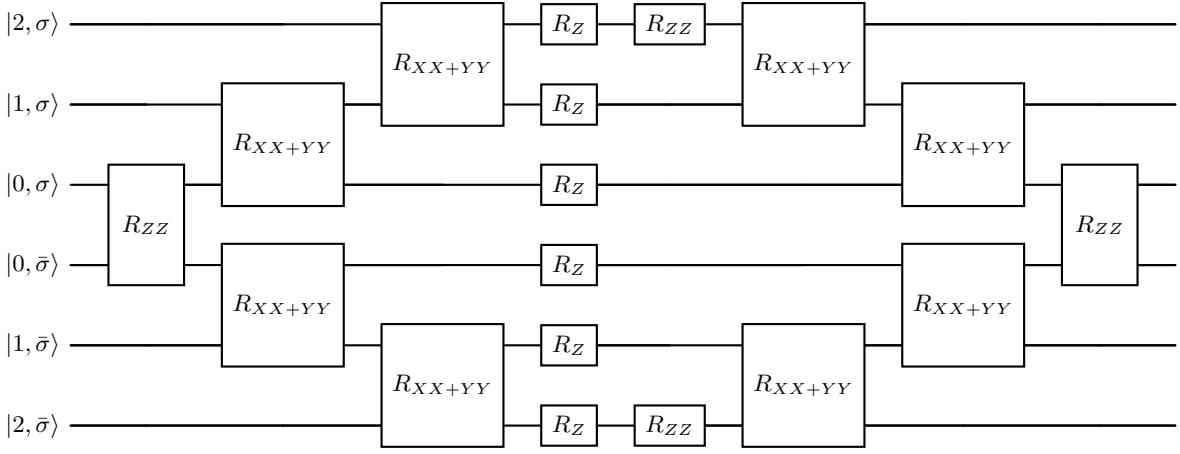

\subsection{Ground state preparation}

For temperature $T=0$, the expectation values in Eq. \ref{timedependetGreensFunctionJWT} are evaluated w.r.t. the ground state $\ket{\text{GS}}$. In our approach, this ground state is approximated by a parametrized circuit ansatz, $\hat{U}(\vec{\theta}_\text{GS})$, which is trained via the cost function,
\begin{equation}
    C(\vec{\theta}_\text{GS}) = \braket{\hat{U}^\dag(\vec{\theta}_\text{GS})|\hat{H}_\text{SIAM}|\hat{U}(\vec{\theta}_\text{GS})}
    \label{costfunctionGroundstate}
\end{equation}
After minimizing the former, the obtained state is proportional to the ground state up to a global phase
\begin{equation}
    \hat{U}(\vec{\theta}_\text{GS})\ket{0}^{\otimes(2B + 2)} \approx e^{i\phi_\text{GS}}\ket{\text{GS}}
\end{equation}
We choose the ansatz to be a Hamiltonian variational ansatz, meaning that a layer in the ansatz is a parametrized version of a Trotter step in a second order Trotterization of the time evolution operator generated by the Hamiltonian in Eq. \ref{SIAMHamiltonian}. 
This ansatz is, as the exact time evolution itself, excitation number conserving for each spin sector. 
Therefore, an initial layer of gates consisting of $R_y$-gates is applied before the actual ansatz, allowing the VQE algorithm to introduce the necessary excitations. 

\begin{figure*}
\centering
\begin{quantikz}[baseline=(current bounding box.center)]
    \lstick[2]{$\sigma$} & \gate{R_Y(\theta_1)} & \gate[wires = 4][1cm]{\hat{U}(\vec{\theta}_\text{GS})} & \qw \\
    \qw & \gate{R_Y(\theta_2)} & \qw & \qw \\
    \lstick[2]{$\Bar{\sigma}$} & \gate{R_Y(\theta_3)} & \qw & \qw \\
    \qw & \gate{R_Y(\theta_4)} & \qw & \qw \\
\end{quantikz}
\Large$\overset{\ket{1110}}{\Rightarrow}$
\begin{quantikz}[baseline=(current bounding box.center)]
    \gate{H} & \ctrl{2} & \gate{\hat{X}} & \gate[wires = 4][1cm]{\hat{U}(\vec{\theta}_\text{GS})}& \qw \\
    \qw & \qw & \gate{\hat{X}} & \qw  & \qw\\
    \qw  & \gate{\hat{X}} & \qw & \qw & \qw\\
    \qw & \qw & \gate{\hat{X}} & \qw  & \qw\\
\end{quantikz}
\caption{VQE circuit for preparing the ground state. After a sufficient number of initial VQE steps, the $R_Y$-layer is replaced by a fixed gate structure depending on the measured computational basis states to introduces the required excitations directly and symmetrically (if it is degenerate) in each spin sector. Afterwards the VQE algorithm is continued until convergency is reached.}
\label{groundstatecircuit}
\end{figure*}
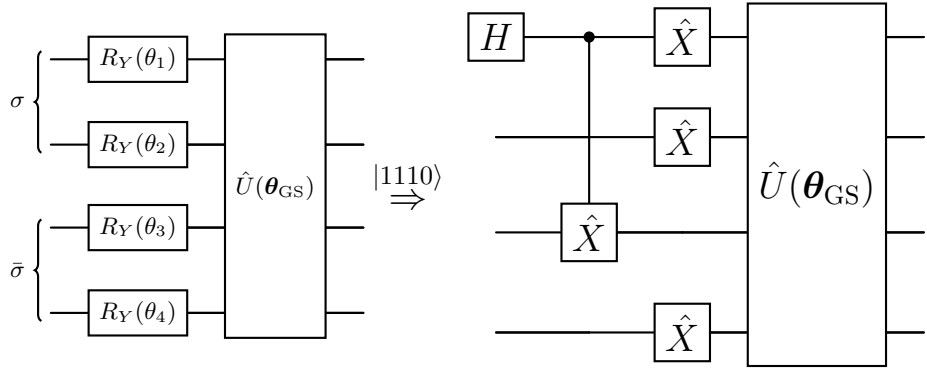

However, by using the Ry-gates the final ground state will also include states with incorrect numbers of excitations due to imperfections in the hardware and to imperfect optimization. 
While these should have a small amplitude, they can still perturb not only the ground state preparation but all the algorithms that follow.
Furthermore, it is possible that the ground state is degenerate in which case one spin-sector has a higher number of excitations than the other. 
The VQE algorithm will most certainly only find one of the two possible ground states. 
To circumvent these issues, we measure the appearing computational basis state after the initial VQE with the $R_y$-gates after some optimization steps.
At this point, the computational basis states with the correct number of excitations will be measured more often than others.
This then allows us to introduce the correct number directly and, in the case of degeneration, symmetrically into the system with a simple combination of Hadamard, $X$- and CNOT-gates thus replacing the layer of $R_y$-gates (see figure \ref{groundstatecircuit}). 
The VQE algorithm is then performed again, using the already found parameters as an initial guess to speed up the process.
If more than one number of excitations is found after the first VQE, this process is repeated and the set up with the lowest energy is used for the ground state.
While this procedure requires more than one VQE step for the ground state, it not only improves the accuracy of the approximation to the ground state energy but also improves the Green's function measurement and the time evolution compression, justifying the additional computational cost.
It further allows the training of the time evolution compression with respect to both ground states simultaneously.
In the following, $\hat{U}_\text{GS}$ refers to the total ground state unitary, i.e. the initial gate layer together with the parametrized ansatz. 

\subsection{Time Evolution Compression}
\label{subsec:timeevolutioncompression}
The circuit depth for the Trotterization of the time evolution increases drastically with the simulation time required for resolving all necessary frequencies in the Fourier transformation of the Green's function.
The implementation of this large number of gates on current hardware is the most dominant challenge for solving the impurity model.
Even though the current two qubit gates have a fidelity over 99\%, the remaining error makes a time evolution with a large number of Trotter steps impossible. 
In recent works, the time evolution is applied with either a low number of Trotter steps, or with a different form of time evolution decomposition. 

Here, we combine the work of ~\cite{Jaderberg,Benedetti,Barison} to compress the time evolution given by a Suzuki Trotter expansion into a parametrized ansatz with fixed depth for each point in time which depends on the system size.
The main idea of the approach is to use a single Trotter steps $\hat{U}_\text{trotter}(\Delta t)$ for the training of the parametrized circuit and compressing the action of many Trotter steps into a shallower circuit. 

Compared to \cite{Jaderberg}, we are not recompiling the full circuit for the Green's function evaluation but only the time evolution to keep the benefits of the ground state preparation discussed above and to reduce the number of parameters to be trained in each iteration. 
We are also not using an adaptive scheme to build our ansatz since that always goes hand in hand with additional costs in resources. 
Furthermore, since the parameters of the SIAM change after each DMFT iteration, it is possible that a once perfect ansatz may not be expressive enough any more, since certain terms are missing. 

As for the ground state preparation, the intuitive choice for an ansatz for the time evolution compression is a Hamiltonian variational ansatz (see appendix \ref{appendix:ansatz} for details). 
Recent work has shown that a Trotterization with a fixed gate count may not be optimal, but its parameters can be optimized to perform the correct time evolution \cite{Mansuroglu}.
This training of the ansatz is performed iteratively, such that a parametrized circuit $\hat{V}(\vec{\theta}_{n})$ for the time step $t_n:= n \Delta t$ is obtained by using the circuit from the previous time step $\hat{V}(\vec{\theta}_{n-1})$ together with a further single Trotter step. That is, one optimizes the parameters $\vec{\theta}_{n}$ such that $\hat{V}(\vec{\theta}_{n})$ best approximates $\hat{U}_\text{trotter}(\Delta t) \hat{V}(\vec{\theta}_{n-1})$.
The optimal approximation should thus lead to a circuit that approximates the entire time evolution,
\begin{equation}
    \hat{V}(\vec{\theta}_{n}) \approx \left(\hat{U}_\text{trotter}(\Delta t)\right)^n.
    \label{optimalApproximation}
\end{equation}
Note that even if the training were done perfectly, the error caused by the Trotterization would remain.
This leads to a trade-off since choosing a smaller $\Delta t$ leads to a larger number of VQA steps to train the ansatz to reach the same point in time.
In future, it would be worthwhile to test whether increasing the number of Trotter steps used in one training step can be used to improve the algorithm.

Moreover, for evaluating the expectation values in the Green's function, Eq. $\ref{optimalApproximation}$ need not be fulfilled for the entire operators. 
It is rather sufficient that our trained ansatz evolves the states $\ket{\text{GS}}$ and $\hat{X}_{0,\sigma}\ket{\text{GS}}$ correctly.
This is accomplished by evolving the states in time with $\hat{V}(\vec{\theta}_n)$, and then evolving them backwards with $\hat{V}^\dag(\vec{\theta}_{n-1})$ and $\hat{U}^\dag_\text{trotter}(\Delta t)$. 
Then, after applying the adjoint of the state preparation for both states, measuring the overlap with $\ket{\vec{0}}$ gives a measure for the quality of the approximation.
While it is possible to train a parametrized circuit for each of the two states individually, this would increase not only the training effort but also the number of gates for calculating the Green's function and the measuring cost.
Training the ansatz to evolve both initial states correctly is therefore favorable, and we apply this strategy. 
Moreover, just as for the VQE of the ground state, a separate training w.r.t. each state would guarantee only the correct time evolution up to a global phase that can differ for both states, i.e.,
\begin{align*}
    \hat{V}(\vec{\theta}_{n})\ket{\text{GS}} &\approx e^{-i\phi_{n_1}} \hat{U}(t_n)\ket{\text{GS}}\\
    \hat{V}(\vec{\theta}_{n})\hat{X}_{1\sigma}\ket{\text{GS}} &\approx e^{-i\phi_{n_2}} \hat{U}(t_n)\hat{X}_{1\sigma} \ket{\text{GS}}.
\end{align*}
In contrast to the ground state preparation, the relative phases for the two states do matter for the Green's function measurement, as the phases do not cancel. For example,
\begin{equation}
\braket{\hat{V}^\dag_{n}\hat{X}_{0,\sigma}\hat{V}_n\hat{X}_{0,\sigma}} \approx e^{-i(\phi_{n_1}-\phi_{n_2})}\braket{\hat{U}^\dag_n\hat{X}_{0,\sigma}\hat{U}_n\hat{X}_{0,\sigma}} \, , 
\end{equation}
where we introduced $\hat{V}_n := \hat{V}(\vec{\theta}_n)$ and $\hat{U}_n := \hat{U}(t_n)$ to simplify the notation.
It is therefore favorable to train the circuit for both states simultaneously, which we do here, ensuring that $e^{-i(\phi_{n_1} - \phi_{n_2})} = 1$ for any $n$.

\subsection{Cost Function}
\begin{figure*}
\centering
\begin{quantikz}
    \lstick{$\ket{0}_\text{anc}$} & \gate{\hat{H}} & \ctrl{1} & \qw  & \ctrl{1} & \gate{\hat{H}} & \meter{} \\
    \lstick{$\ket{0}^{\otimes 2B + 2}$} & \gate{\hat{U}_\text{GS}} & \gate{\hat{X}_{0,\sigma}} & \gate{\hat{L}_n} & \gate{\hat{X}_{0,\sigma}} & \gate{\hat{U}^\dag_\text{GS}} & \meter{} 
\end{quantikz}
\caption{Circuit for measuring the global and local versions of the cost function}
\label{measurementGlobalCostFunction}
\end{figure*}
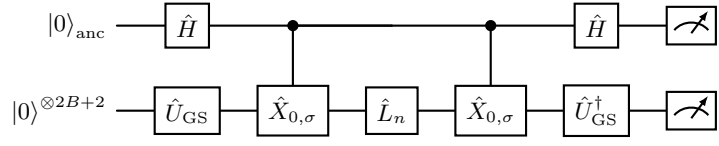

To simplify the notation, we introduce the two training circuits for the parameters $\vec{\theta}_n$,
\begin{align}
    L_n &:= \hat{V}^\dag(\vec{\theta}_{n-1})\hat{U}^\dag_\text{trotter}(\Delta t) \hat{V}(\vec{\theta}_n)\\
    K_n &:= \hat{X}_{0,\sigma} L_n \hat{X}_{0,\sigma} \, ,
\end{align}
and define the cost function to be
\begin{equation}
    C(\vec{\theta}_n) = 1 - \text{Re}\left(\braket{\text{GS}|L_n^\dag|\text{GS}}\braket{\text{GS}|K_n|\text{GS}}\right)
    \label{globalCostFunction}
\end{equation}
In appendix \ref{appendix:costfunction} we show that this cost function in fact returns the correct time evolution approximations while also setting the phase factor to unity, $e^{-i(\phi_{n_1} - \phi_{n_2})} = 1$.
To evaluate $\text{Re}\left(\braket{\text{GS}|L_n^\dag|\text{GS}}\braket{\text{GS}|K_n|\text{GS}}\right)$ we run the circuit shown in figure \ref{measurementGlobalCostFunction}, where measuring the observable $O_\text{global}:=\hat{Z}_\text{anc}\otimes\left(\ket{0}\bra{0}\right)^{\otimes 2B +2}$ returns the required term. 

Global cost functions, as in Eq. \ref{globalCostFunction} have been shown to suffer from Barren plateaus (for hardware efficient ansatzes), even for shallow (hardware efficient ansaetze) \cite{Cerezo2021}. 
By measuring $O_\text{local}:= \hat{Z}_\text{anc} \otimes \left(\ket{0}\bra{0}\right)_p$ for each qubit $p$ instead of $O_\text{global}$ the cost function can be transformed into an equivalent local version:
\begin{equation}
    C(\vec{\theta}_n) = \sum_{p,\sigma} \left(1 - \text{Re}(\braket{\text{GS}|L_n^\dag\hat{U}_\text{GS})|0}\braket{0|_{p,\sigma} \hat{U}^\dag_\text{GS}K_n|\text{GS}})\right)
    \label{localCostFunction}
\end{equation}
We can further rewrite the observable $O_\text{local}$ as $\left(\ket{0}\bra{0}\right)_{p,\sigma} = \frac{1}{2}\left(\hat{I}_{p,\sigma}+\hat{Z}_{p,\sigma}\right)$ such that
\begin{align}
    C(\vec{\theta}_n) & = (B+1)\left(1 - \text{Re}(\braket{\text{GS}|L_n^\dag K_n|\text{GS}})\right)  \label{bottomCostTerms} \\
    &+ \sum_{p,\sigma} \frac{1}{2}\left(1-\text{Re}(\braket{\text{GS}|L_n^\dag\hat{U}_\text{GS}\hat{Z}_{p,\sigma}\hat{U}^\dag_\text{GS} K_n|\text{GS}})\right) \nonumber
\end{align}
The terms in Eq. \ref{bottomCostTerms} can be further simplified. 
For this we look at the optimized ground state preparation circuit. 
The operator $\sum_iZ_i$ commutes with the Hamiltonian variational ansatz used in the ground state preparation, but it does not with the gates introducing initial excitations into the system (see Appendix \ref{appendix:costfunctionreduction} for details). 
Assuming a spin-symmetric number of excitations (i.e., the initial gates are just $\hat{X}-$gates applied to a set of qubits $M_\text{ini}$), we can use the relations $\hat{X}\left[\hat{Z},\hat{X}\right] = -2\hat{Z}$ and $\hat{X} \hat{Z} \hat{X} = -\hat{Z}$ to write
\begin{align*}
    &\sum_{p,\sigma} \left(1-\text{Re}(\braket{\text{GS}|L_n^\dag\hat{U}_\text{gs}\hat{Z}_{p,\sigma}\hat{U}_\text{GS}^\dag K_n|\text{GS}} \right)\\ 
    &= 2(B+1) - \text{Re}(\braket{\text{GS}|L_n^\dag\hat{U}_\text{GS}\sum_{p,\sigma}\hat{Z}_{p,\sigma}\hat{U}_\text{GS}^\dag K_n|\text{GS}} \\
    &= 2(B+1) - \sum_{j\in M_\text{ini}}2\text{Re}(\braket{\text{GS}|L_n^\dag\hat{U}_\text{GS}\hat{Z}_j\hat{U}_\text{GS}^\dag K_n|\text{GS}}) \\
    &- \sum_{p,\sigma} \text{Re}(\braket{\text{GS}|L_n^\dag\hat{Z}_{p,\sigma} K_n|\text{GS}}
\end{align*}
The cost function is therefore given by
\begin{align}
    C(\vec{\theta}_n) & = (B+1)\left(1 - \text{Re}(\braket{\text{GS}|L_n^\dag K_n|\text{GS}})\right)\label{finalCostAnc}\\
    &+ (B+1) - |M_\text{ini}| - \frac{1}{2}\sum_{p,\sigma} \text{Re}(\braket{\text{GS}|L_n^\dag\hat{Z}_{p,\sigma} K_n|\text{GS}}    \label{finalCostZAnc}\\
    &+ |M_\text{ini}| - \sum_{j\in M_\text{ini}}\text{Re}(\braket{\text{GS}|L_n^\dag\hat{U}_\text{GS}\hat{Z}_j\hat{U}_\text{GS}^\dag K_n|\text{GS}}) 
    \label{finalCostUgsZAnc}
\end{align}
In the case where $B+1 = |M_\text{ini}|$, the term $\sum_{p,\sigma} \text{Re}(\braket{\text{GS}|L_n^\dag\hat{Z}_{p,\sigma} K_n|\text{GS}}$ must vanish. 
Since our ansatz conserves the number of excitations, this is however known a prior and hence the term need not be measured. 
We therefore split the cost functions into terms that require a second application of $\hat{U}_\text{GS}$, as in Eq. \ref{finalCostUgsZAnc}, and terms that do not, as in Eqs. \ref{finalCostAnc} and \ref{finalCostZAnc}. 
By doing so, the error during training caused by imperfect gates can be reduced for the first two terms. 
Further, we note that each measurement used for evaluating the terms in Eq. \ref{finalCostUgsZAnc} can also be used for evaluating the terms in \ref{finalCostAnc}.
For optimizing the cost, we use a gradient based method. 
Evaluating the gradient of \ref{localCostFunction} is done by using the parameter shift rule (see Appendix \ref{appendix:parametershiftrule} for details). 

There are modifications of the cost function which would require fewer gates when run on real hardware but do not affect the numerical simulations of the algorithm, that we consider here. Nonetheless we discuss these briefly to prepare for possible future implementations on real hardware.

\subsection{Reducing gate count in the cost function}
Evaluating the cost term would still require the second application of the gates for the ground state preparation, thus increasing the circuit depth of the training circuit. This can be circumvented by using the following cost function instead: 
\begin{align*}
    C(\vec{\theta_n}) &= \left(1 - \text{Re}(\braket{\text{GS}|L_n^\dag K_n|\text{GS}})\right) \\
    &+ \text{Re}(\braket{\text{GS}|L_n^\dag \hat{H}_\text{SIAM} K_n|\text{GS}})
\end{align*}
Here, $\hat{h}_i$ are the terms in the Hamiltonian that are measured in combination of the $\hat{Z}$ expectation value of the ancilla qubit.
The first term in this cost function guarantees that $K_n\ket{\text{GS}} = L_n\ket{\text{GS}}$ while the energy measurement leads to $K_n\ket{\text{GS}} = \ket{\text{GS}}$ up to a global phase. 
The Hamiltonian mostly consists of terms that can be measured in the Z-basis. 
By increasing the number of bath sites, additional terms of the form $\hat{X}\hat{Z}\dots\hat{Z}\hat{X}$ must be measured, where the number of measurement setups increases only linearly with the number of bath sites. 
In addition, each measurement setup can also be used as a measurement for the initial term that measures the overlap. Alternative constructions that do not require the application of $\hat{U}^\dag_\text{GS}$ are possible, but these require a number of measurements that is factorial in the number of excitations and are therefore less optimal.
While this cost function could potentially improve results of the training of the parametrized circuits on hardware at the cost of more measurement setups, it does not provide improvements in regards of simulations and is therefore not used for our results.

\subsection{Remarks regarding the Ansatz}
For designing the ansatz, one has to consider, that the final gates of the ansatz, which commute with each other and with the controlled $X/Y_{0,\sigma}$-gate required for measuring $G(t)$, cancel in the evaluation of the Green's function. 
Hence, their contribution to $G(t)$ vanishes. 
Yet the significance of these gates can be higher compared to their counterparts in a Trotter expansion because of the compression and may lead to inaccurate values of the Green's function.
To avoid this issue, the Trotter expansion and therefore the ansatz are chosen as shown in figure \ref{ansatz}.
Note that in this way, none of the gates inside the cone commutes with the final $R_{ZZ}$-gate or the gates following it.
This also allows evaluating the local cost function \ref{localCostFunction} without losing the information of too many other gates in the first place.
Furthermore, the cone shape guarantees that all gates are efficiently trained, see \cite{Barison}.

While further optimizations of the ansatz are possible for specific cases, we have chosen to use an ansatz that works generically, such that it can be used in every DMFT-iteration without changing the gate structure from iteration to iteration. 
This also allows using the parameters of the previous iteration as the starting point for the current VQA step since the SIAM parameters do not drastically change between iterations (after the initial steps).
The cost function is therefore already close to the minimum, guaranteeing a warm start and making the algorithm more resilient to Barren plateaus.
In addition, we can easily compare the gate count between different SIAM parameters and between systems with different numbers of bath sites. 
The ansatz is also designed to work on a linear chain of qubits on a superconducting hardware architecture. 
The FSWAP-gates, required for turning the hybridization terms into two-qubit gates, are incorporated into the $R_{XX+YY}$-gates, and are therefore do not increase the two-qubit gate count. In total, a layer of our ansatz consists of $2+4B$ two qubit gates, where $B$ is the number of bath sites.

\subsection{Classical Postprocessing of $G(t)$}
After obtaining an approximation of the time evolution the Greens function can be evaluated on the quantum device, as shown in figure \ref{GFmeasure}, by replacing $\hat{U}(t_n)$ with our approximation $\hat{V}(\vec{\theta}_n)$. 
The iterative scheme for DMFT requires a Fourier transformation of the time dependent Green's function to get $G(\omega)$. 
However, as already stated before, performing the time evolution is either restricted by the number of Trotter steps that can be used on current hardware or, as in the case here, by the number of compression steps one must perform to obtain its approximation. 
Instead of just performing the transformation, we thus use an ansatz for the Lehmann representation of the Green's function,
\begin{align}
    iG^R_{\text{imp},\sigma}(t) &= \sum_j |\braket{j|\hat{c}^\dag_{i,\sigma}|\text{GS}}|^2e^{-i(E_j - E_\text{GS})t} \\
    &+ |\braket{j|\hat{c}_{i,\sigma}|\text{GS}}|^2 e^{i(E_j-E_\text{GS})t}\\
    &:= \sum_j \alpha_je^{-i\omega_jt} + \beta_je^{i\omega_jt}
\end{align}
and fit it to $G(t)$. 
The so obtained Lehmann parameters, $\alpha_j$, $\beta_j$ and $\omega_j$ can then be directly used to compute the Green's function in the frequency domain,
\begin{equation}
    G_\text{imp}(\omega+i\eta) = \sum_j\frac{\alpha_j}{\omega + i\eta-\omega_j} + \frac{\beta_j}{\omega+i\eta+\omega_j}.
\end{equation}
leading to better result for the DMFT-loop than the Fourier Transformation. 
Nevertheless, we use a Fast-Fourier-Transformation (FFT) to get a good initial guess for some of the Lehmann parameters. As this, however, does not capture every frequency well, especially for small $\alpha_j/\beta_j$, we  use the conditions \cite{Rungger},
\begin{align}
    \sum_j \alpha_j + \beta_j &= 1\label{LPconditionSum}\\
    G(\omega = \epsilon_p) &= 0 \quad \forall p\label{LPconditionRoot}\\
    \left.\frac{\partial G(\omega)}{\partial \omega}\right|_{\omega = \epsilon_p} &= \frac{1}{V_p^2} \quad \forall p\label{LPconditionDeriv}
\end{align}
to find the remaining parameters (see appendix \ref{appendix:greensfit} for details). 
In total, the scheme to find the Lehmann parameters is as follows:
\begin{enumerate}
    \item Fourier transform $G(t)$ to get the some of the $\omega_j$
    \item Use $\omega_j$ as an initial guess to find their respective $\alpha_j$ and $\beta_j$. Here the conditions \ref{LPconditionSum} - \ref{LPconditionDeriv} have already been applied.
    \item If $\sum_j \alpha_j + \beta_j < 1$, add an additional parameter to the fitting procedure until $|\sum_j \alpha_j + \beta_j - 1|\leq \delta$ for some threshold $\delta$ or no further improvement can be expected. 
\end{enumerate}
During step 3.) the already found Lehmann parameters are allowed to adjust in a small range around their values during the fitting.
Recently, another method for analyzing real-time data of quantum systems \cite{Gull2025} has been proposed.
This approach works well for a continuous spectra, but in the case of a small number of bath sites, the spectra is discrete.
Nevertheless, since the number of poles is exponential, the spectra should basically become continuous if the number of bath sites is further increased so that this method could prove to be useful here as well.

\section{Results}
We verified the performance of our approach in numerical experiments using the cost function \ref{finalCostAnc} - \ref{finalCostUgsZAnc}. To this end, we investigated all its crucial components and present the results in this section. We start with the assessment of the accuracy of the  preparation of the ground state of the SIAM. Then we present results that show the convergence of the DMFT iteration. Finally we discuss the achievable compression of the time evolution circuits for the time dependent Greens functions.

\subsection{Ground state preparation}
\begin{figure*}[t]
    \centering
    \begin{minipage}{0.49\textwidth}
        \includegraphics[width=0.9\linewidth]{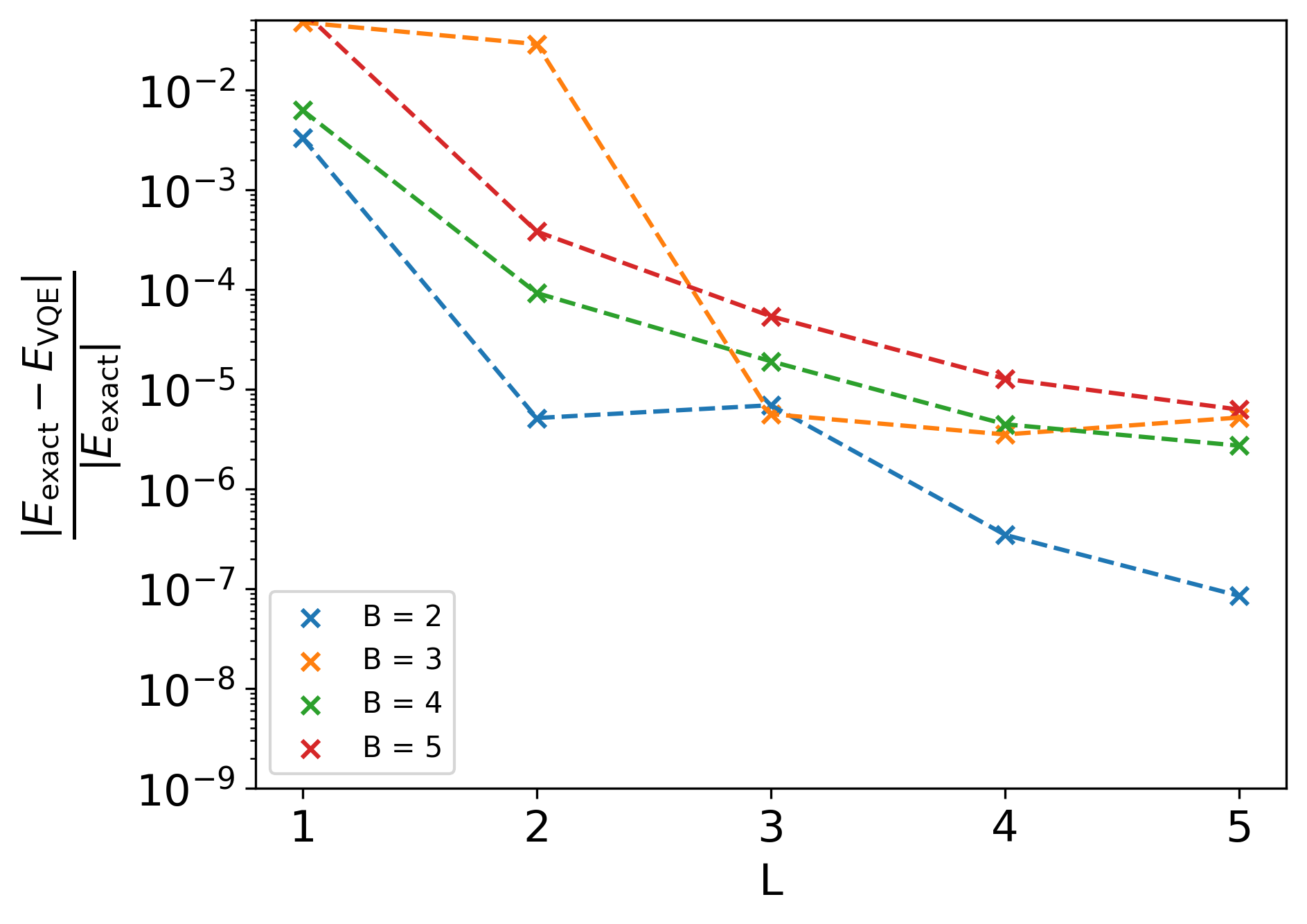}
    \end{minipage}
    \begin{minipage}{0.49\textwidth}
        \includegraphics[width=0.9\linewidth]{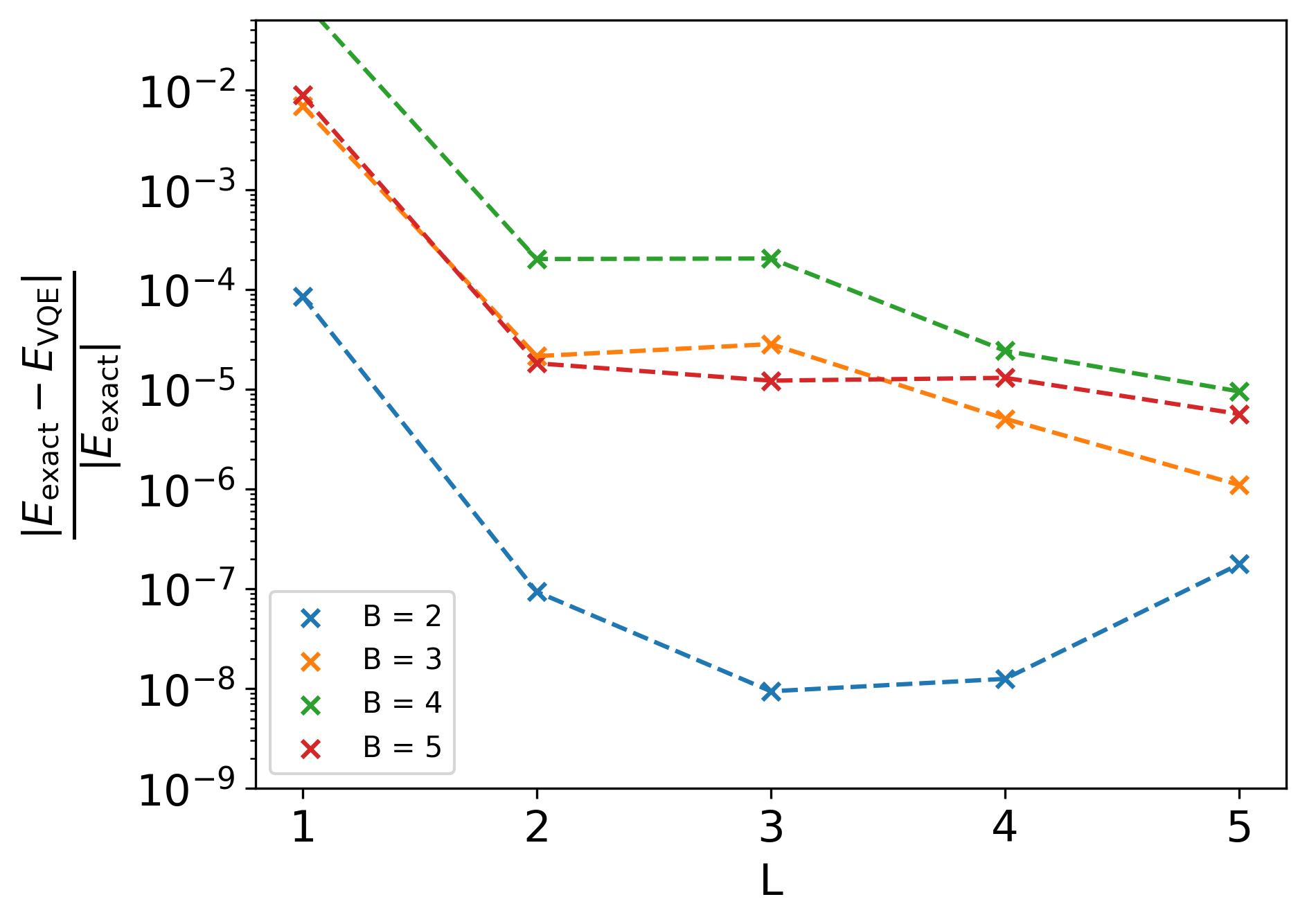}
    \end{minipage}
    \caption{Error in the ground state energy for $U = 4v$ for different fillings ($n=1$ left, $n = 0.5$ right). $L$ denotes the number of layers in the ansatz. In each case at most $L \leq B$ layers are necessary for an error below $10^{-4}$.}
    \label{gsResults}
\end{figure*}

Our scheme of encoding the excitations directly and using an excitation number conserving ansatz allows for finding the ground state energy with high accuracy in simulations, where the accuracy only depends on the number of layers used.
This number is directly related to the number of bath sites, $B$, and does not depend on the parameters of the SIAM.
We found that for an error of $\frac{|E_\text{exact}- E_\text{VQE}|}{|E_\text{exact}|}<10^{-4}$ at most $B$ layers are required, see figure \ref{gsResults} for explicit values of $B$.
With a total two-qubit gate count of $2+4B$ for each layer, the required number of two-qubit gates scales quadratically with the number of bath sites. 

The results shown in figure~\ref{gsResults} are obtained using the SIAM parameters after the DMFT loop has converged.
The behavior of the algorithm for these parameters is representative for most of the DMFT iterations. 
During the DMFT loop, the bath parameters vary only by a relatively small amount, so that the number of required layers is not affected. 
Only for the initial iterations, where bath parameters are given by a simple initial guess, we find that fewer layers are necessary for the ground state preparation.
The data points $B=3, L=2$ at half-filling and for $B=4, L=5$ away from half-filling show a larger error because the ground state preparation converged to a state with incorrect excitation number. 

\subsection{DMFT Loop Convergence}
\label{subsec:dmftloopconvergence}
\begin{figure}
    \centering
    \includegraphics[width = 0.9\linewidth]{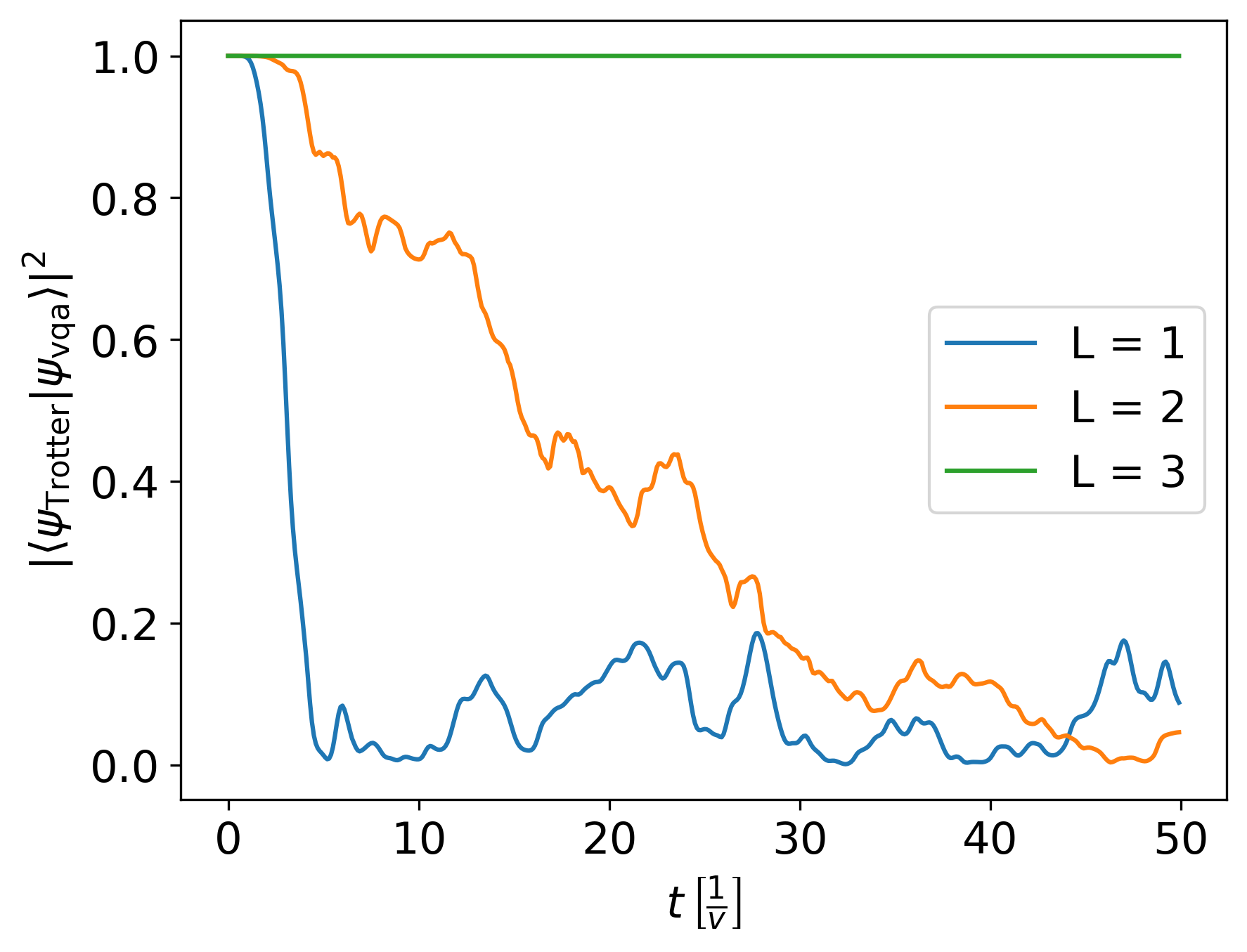}
    \caption{Fidelity of the trained compressed time evolution applied to $\hat{X}_{0,\sigma}\ket{\text{GS}}$ compared to the same time evolution using a second order Trotterization with $\Delta t= 0.1\frac{1}{v}$}.
    \label{fidelityDMFTIteration}
\end{figure}

For exploring the convergence of the DMFT iteration loop, we used the time evolution compression, introduced in section \ref{subsec:timeevolutioncompression}, for a DMFT-Loop for the Hubbard model with $U=4v$ and $n=0.5$. 
For illustration purposes, we here only consider the case with two bath sites but generalizations to larger systems are straight forward.
We simulated the system up to $t_\text{max} = 50\frac{1}{v}$ and used a Trotter step size of $\Delta t = 0.1\frac{1}{v}$. 
Here we illustrate the simulated (noise free) results obtained for a single iteration in the middle of the DMFT loop.
As shown before, the ground state preparation requires two layers of gates for this case. For the compressed time-evolution, we instead need three layers (which did not change with iteration during the DMFT loop).
Using fewer layers allows for an error free time evolution (compared to the Trotterization) only up to a certain (small) point in time (see figure \ref{fidelityDMFTIteration}).
For $L=3$ layers, instead, we can reach arbitrary long times (note that the longest times in figure~\ref{fidelityDMFTIteration} corresponds to $500$ Trotter steps). This suggests that  with an efficient and accurate training, $G(t)$ can be evaluated for any time scale without increasing the circuit depth. 
The  number of layers that is required for a perfect evolution over long times will thereby still depend on system size, which will be analysed below. We illustrate the optimization process in appendix \ref{appendix:qcresources}.

\begin{figure}
    \centering
    \includegraphics[width = 0.9\linewidth]{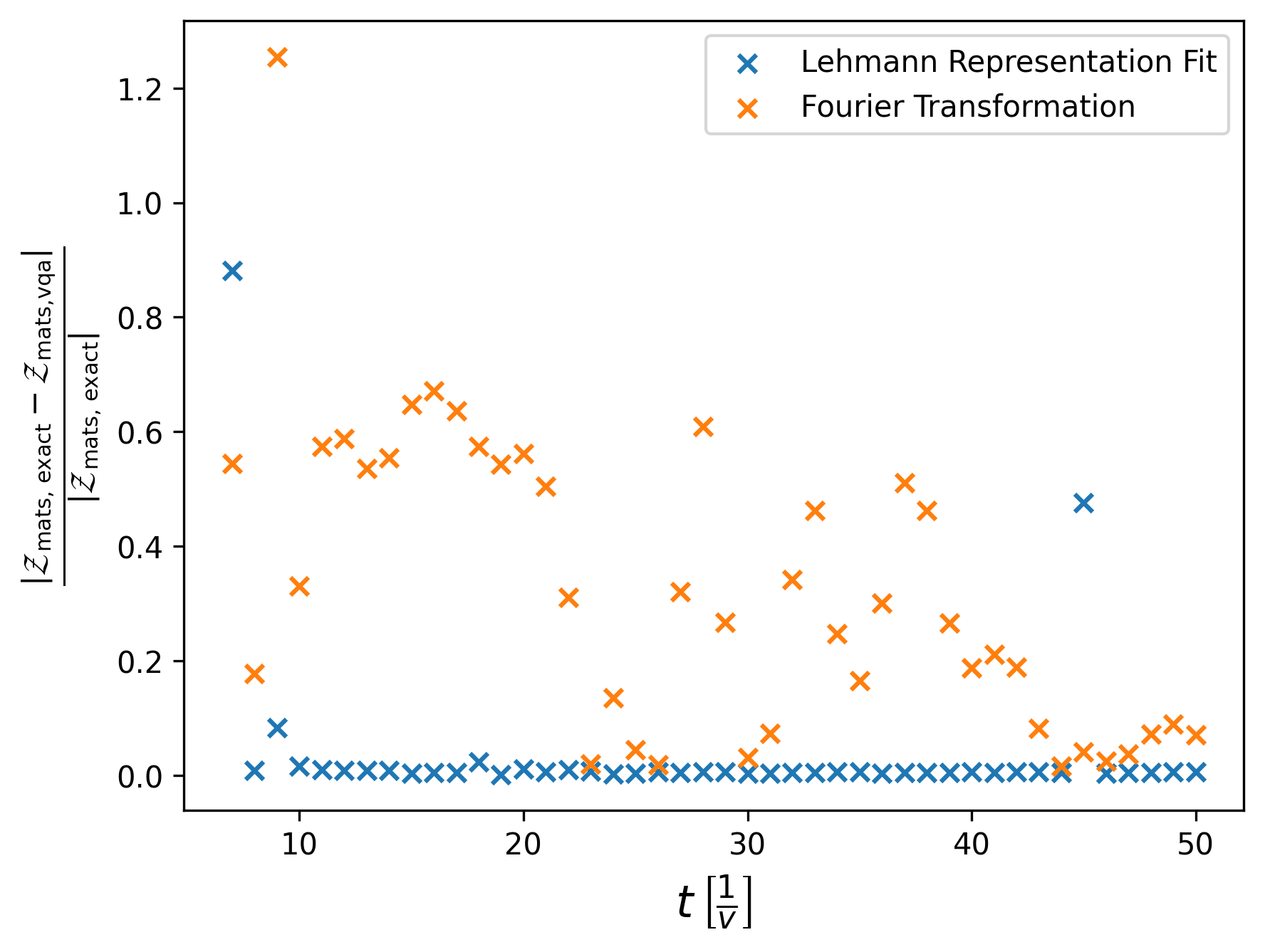}
    \caption{$\mathcal{Z}_\text{mats}$ obtained using Fourier transformation and the fitting procedure for different times. Here we used $\eta = 0.1$ and a finite temperature of $\beta = 200$. Using the fitting procedure a stable value of the quasiparticle weight can be found starting at $t = 10 \frac{1}{v}$ while the result obtained from the Fourier transformation is not stable.}
    \label{qpwDMFTIteration}
\end{figure}

\begin{figure*}
    \centering
    \begin{minipage}{0.49\textwidth}
        \includegraphics[width=0.9\linewidth]{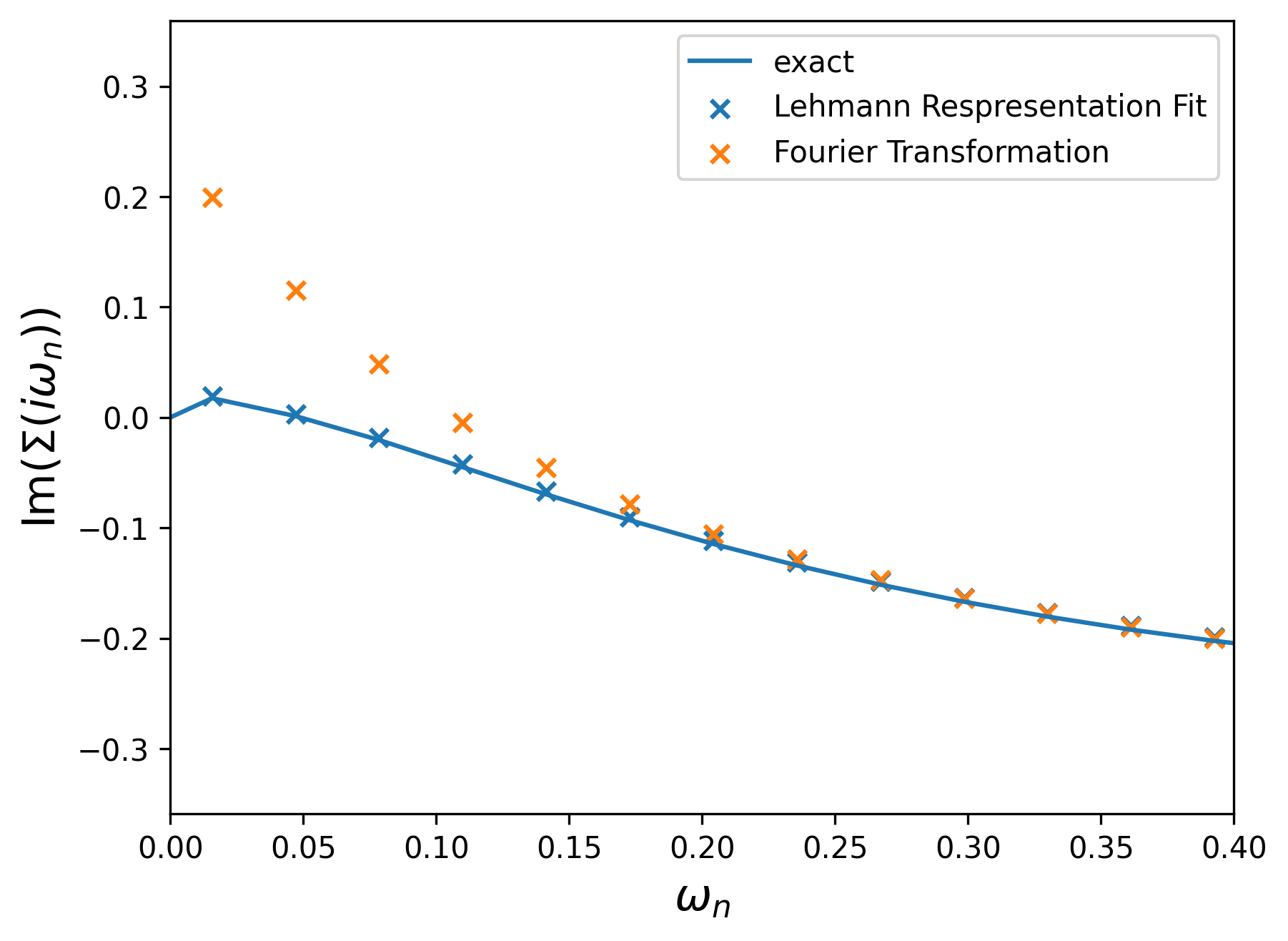}
    \end{minipage}
    \begin{minipage}{0.49\textwidth}
        \includegraphics[width=0.9\linewidth]{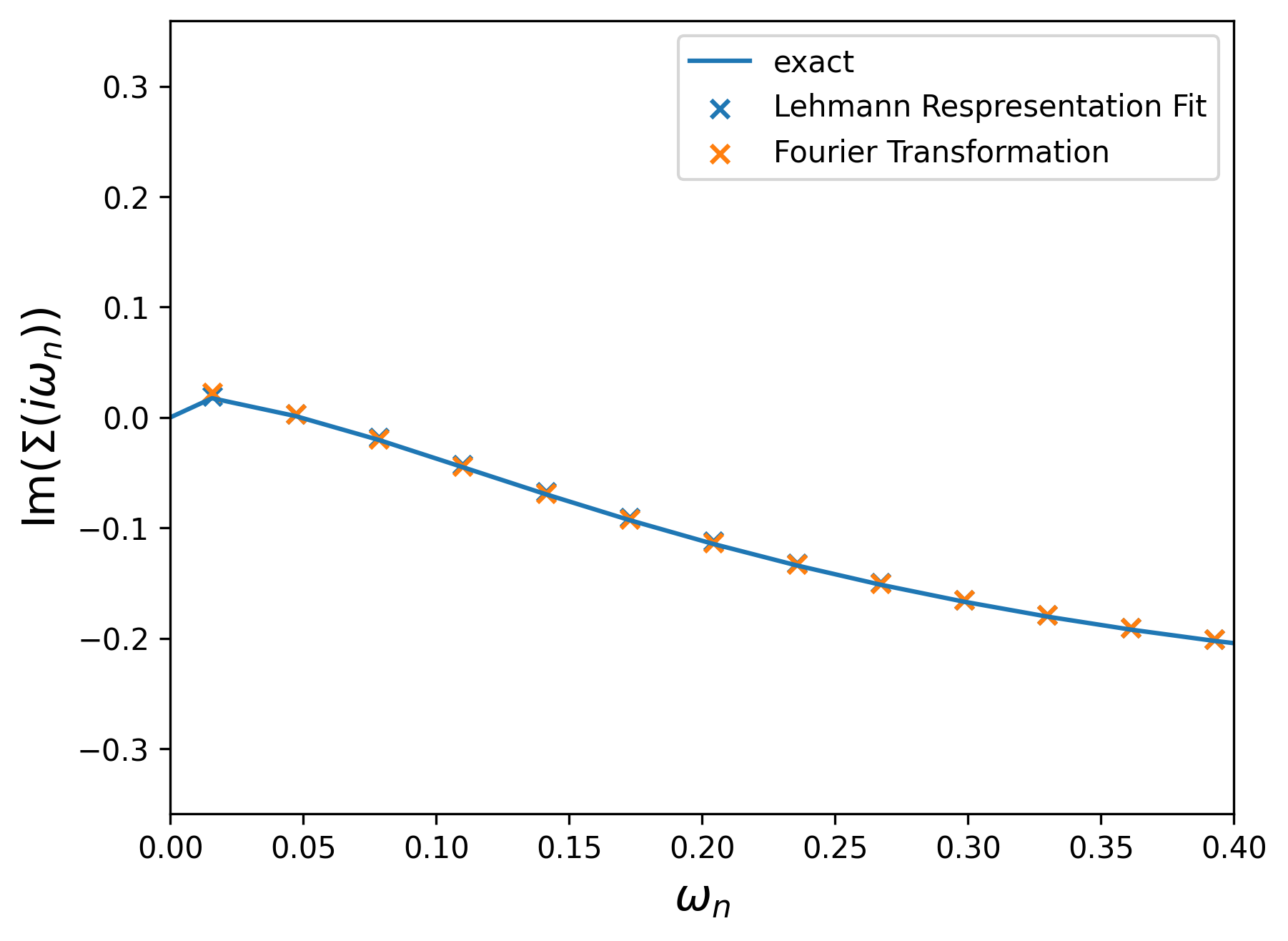}
    \end{minipage}
    \caption{Matsubara self energy $\Sigma(i\omega_n)$ ($\eta = 0.1$, $\beta = 200$ obtained using the fitting procedure and the Fourier transformation applied to $G_\text{imp}(t)$ for the time range up to $t=12\frac{1}{v}$ (left) and $t=50\frac{1}{v}$. The fitting procedure provides a better results using a smaller time range.
    } 
    \label{matsubaraSE}
\end{figure*}

Nevertheless, noisy quantum devices render this very challenging as errors in training accumulate. 
Using the presented fitting procedure, the required time scale can however be shortened drastically. 
As seen in figure \ref{qpwDMFTIteration}, the quasiparticle weight can be evaluated for times much shorter than with the Fourier transformation alone. 
The latter was also found to be unstable, even without the addition of noise. 
This can also been seen in the Matsubara self-energy $\Sigma(i\omega_n)$ for $\omega_n\rightarrow0$, i. e. in the region which is important for calculating the quasiparticle weight.
For shorter times, the fitting procedure already provides a good solution, while the Fourier transformation only fits the dynamics well for higher frequencies (see figure \ref{matsubaraSE}).

\subsection{Long Scale Time Evolution Compression}
\begin{figure*}
    \centering
    \begin{minipage}{0.49\textwidth}
        \includegraphics[width=0.9\linewidth]{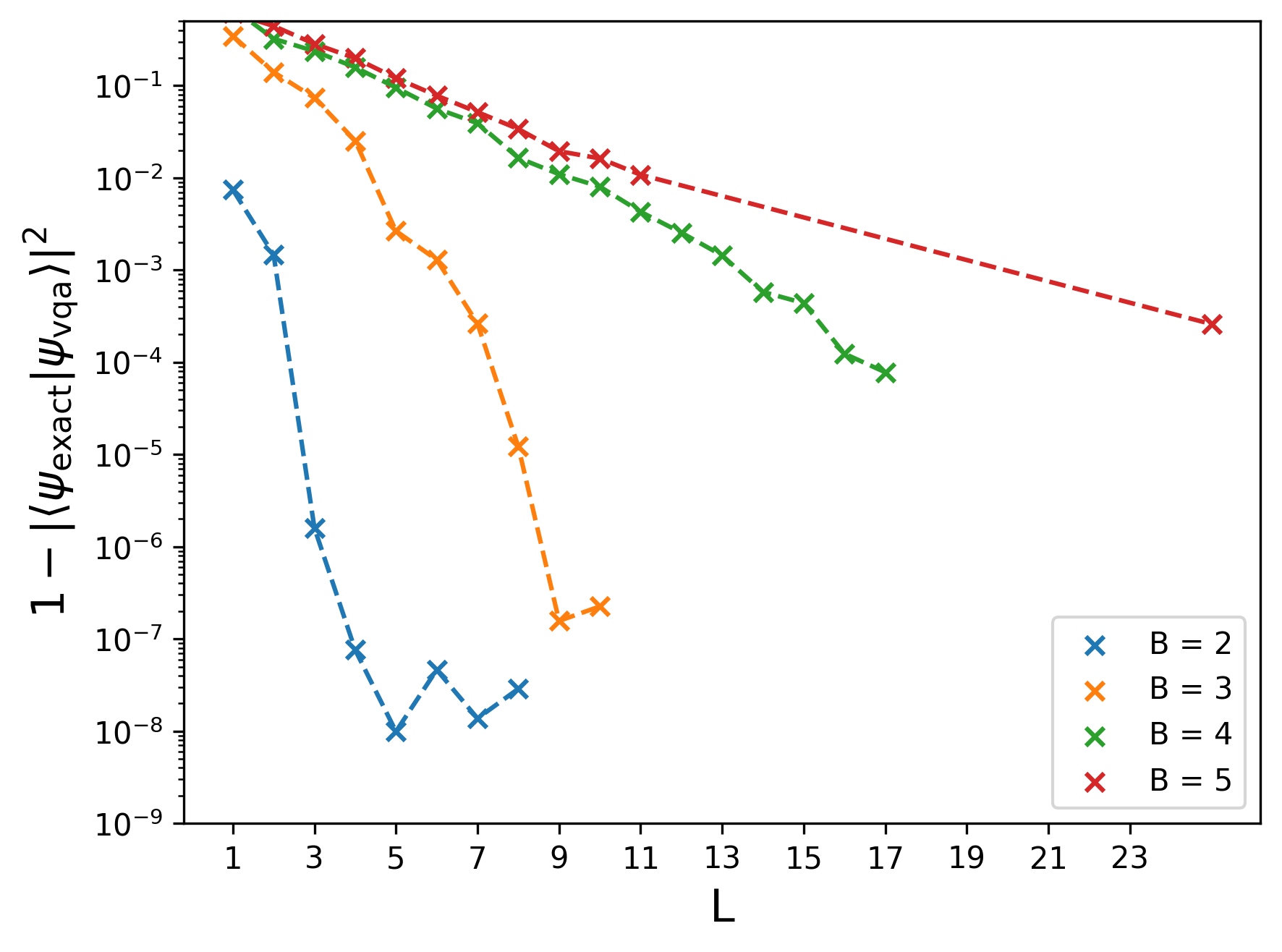}
    \end{minipage}
    \begin{minipage}{0.49\textwidth}
        \includegraphics[width=0.9\linewidth]{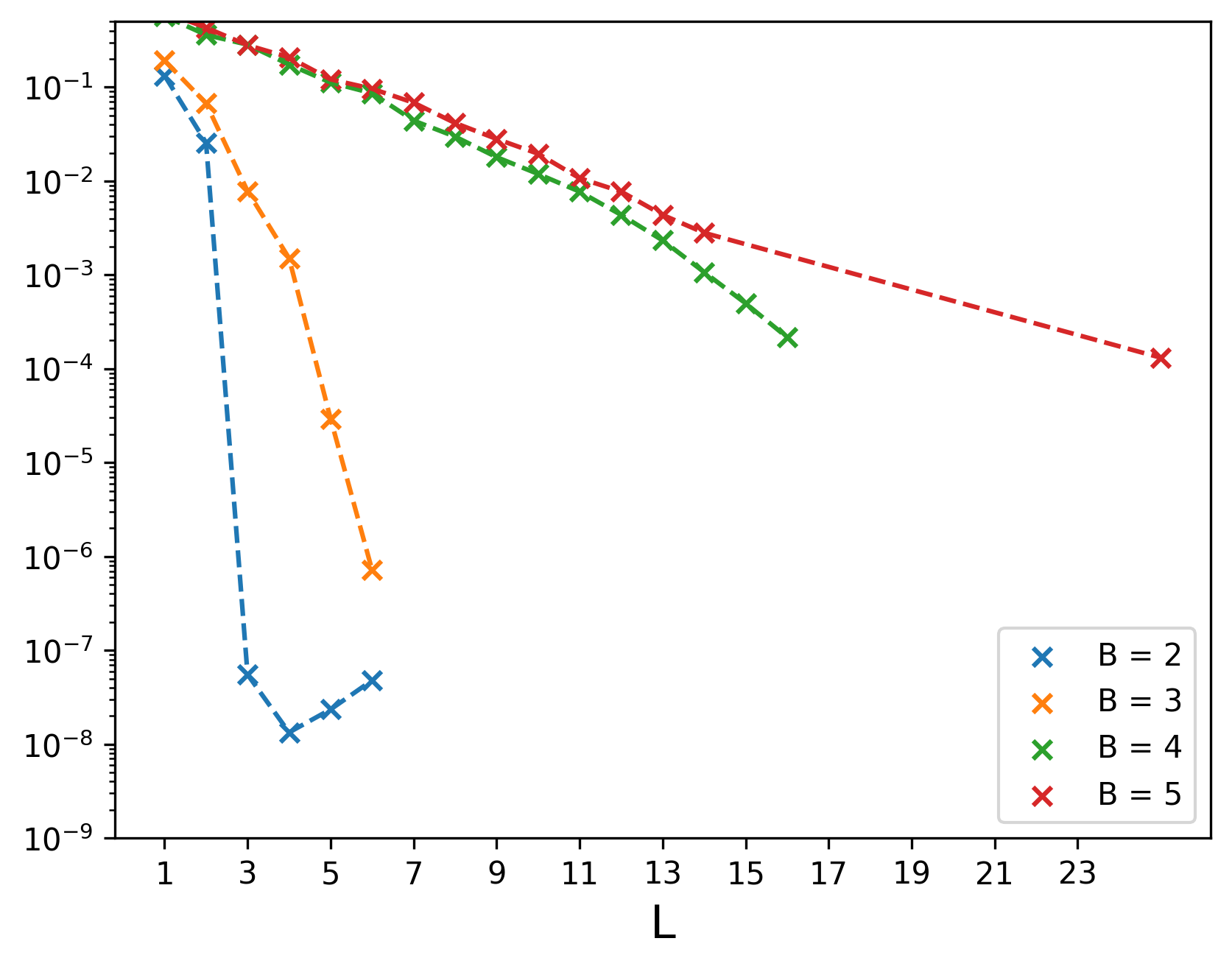}
    \end{minipage}
    \caption{Infidelity of the approximated and exact time evolution as a function of the number of layers $L$ in the ansatz for $U = 4v$, $t = 1000\frac{1}{v}$ and with different fillings ($n=1$ left, $n = 0.5$ right). The overlap is calculated using the state $\hat{X}_{0,\sigma}\ket{\text{GS}}$ (in case of degeneracy a single ground state). In each case at most $L \leq B^2$ layers are necessary for an error below $10^{-4}$.}
    \label{tcResults}
\end{figure*}

\begin{figure*}
    \centering
    \begin{minipage}{0.49\textwidth}
        \includegraphics[width=0.9\linewidth]{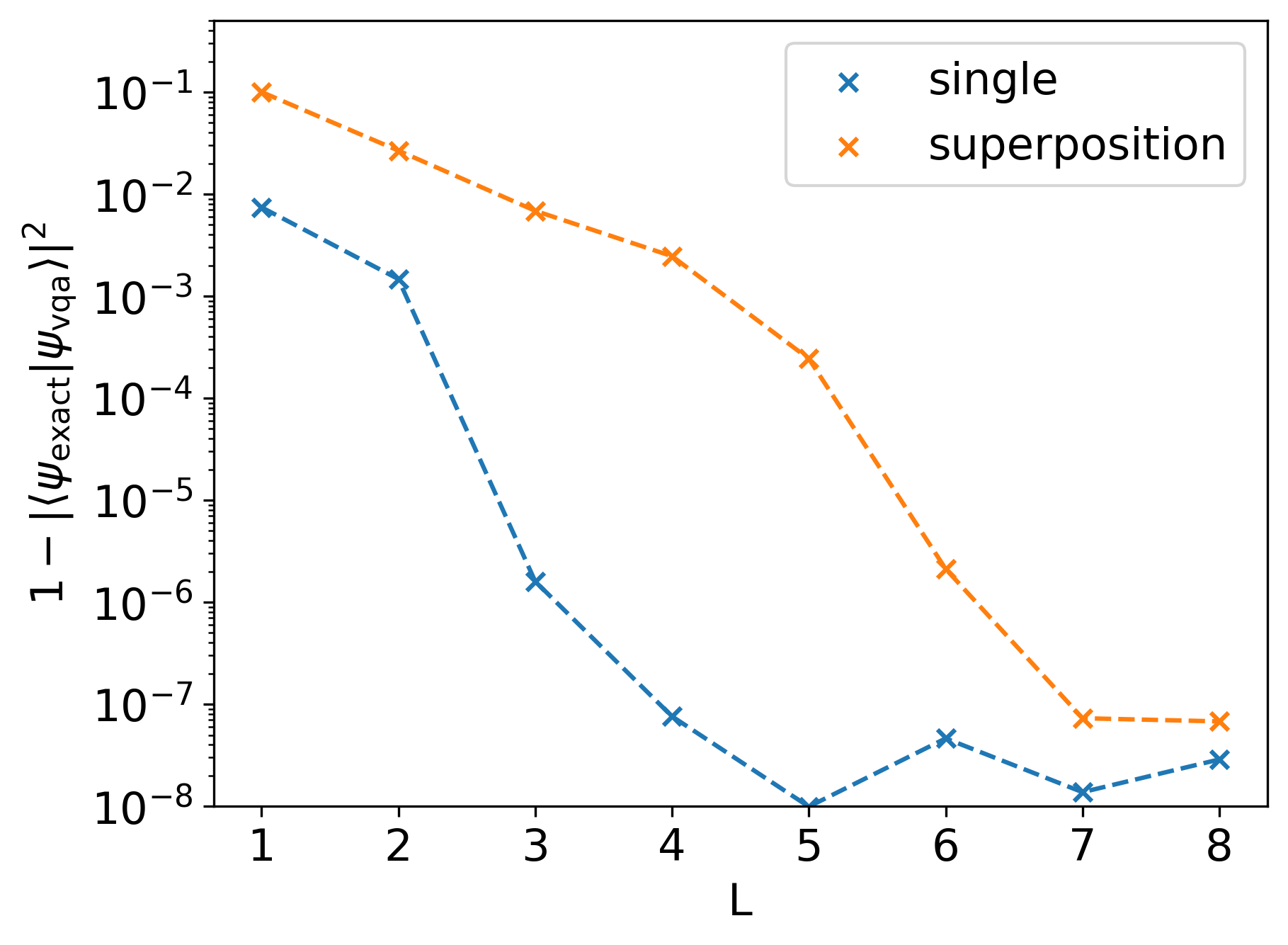}
    \end{minipage}
    \begin{minipage}{0.49\textwidth}
        \includegraphics[width=0.9\linewidth]{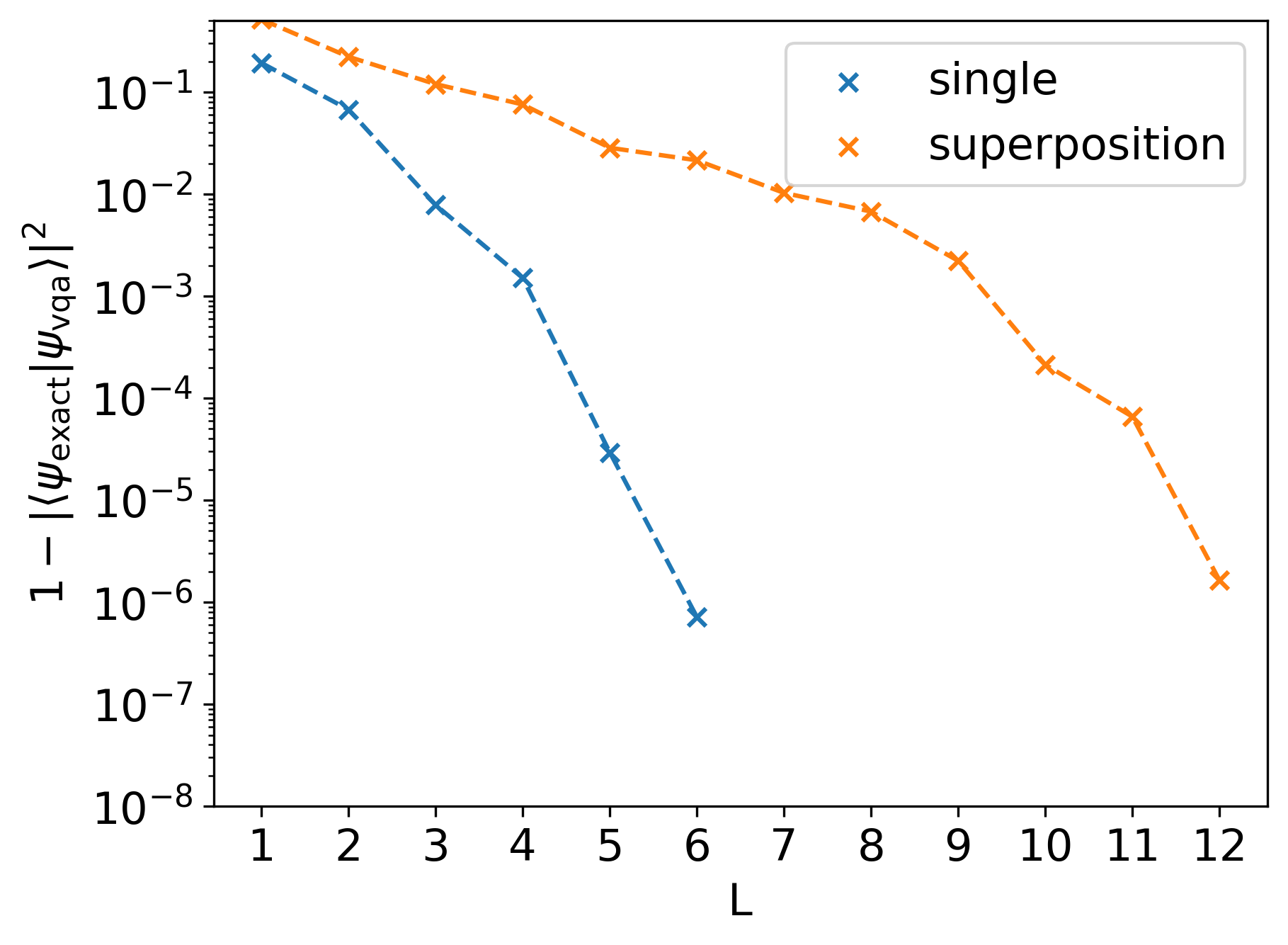}
    \end{minipage}
    \caption{Infidelity of the approximated and exact time evolution  as a function of the number of layers $L$ in the ansatz for $U = 4v$, $t = 1000\frac{1}{v}$ ($B = 2$ and $n=1$ left, $B=3$ and $n = 0.5$ right). The overlap is calculated using the one single state $\hat{X}_{0,\sigma}\ket{\text{GS}}$ or the superposition of the both. For the superposition, twice as many layers are required to reach an error below $10^{-4}$.}
    \label{tcResultsSuperPosition}
\end{figure*}

Small simulation times can be reached with a small number of layers, but at some point the fidelity with the actual time evolution drops, see section \ref{subsec:dmftloopconvergence}. 
This is not caused by the accumulation of errors during the optimization (at least not in the exact simulation) but rather by the lack of expressiveness of the ansatz for the time evolution.
To verify this, we train the ansatz with different numbers of layers by utilizing the exact time evolution for a long simulation time (here $t = 1000\frac{1}{v})$.
The infidelity of the approximation to the exact time evolution, both evaluated w. r. t. the state $\hat{X}_{0,\sigma}\ket{\text{GS}}$ (in case of degeneracy, only a single state), is illustrated in figure \ref{tcResults} for different numbers of bath sites and fillings. 
As for our results for the ground state preparation, here we also used the SIAM parameters after the DMFT loop converged. 
In general, the infidelity drops below $10^{-4}$ if the ansatz consists of at least $B^2$ layers.
In case of the degeneracy, using the superposition of the two ground states instead of a single ground state increases the required layers by a factor of two to achieve the same fidelity (see figure \ref{tcResultsSuperPosition}).

\section{Discussion}
In this paper, we presented a variational time evolution compression algorithm for specifically solving the single impurity Anderson model on a quantum device.
Most earlier works considered a single specific case of the Hubbard model with half filling and choose for the SIAM an approximation with just one bath site (two-site DMFT \cite{Potthoff2001}).
In contrast to this, we did not restrict the application of our algorithm to these simplifications, thus showing its potential for more general and thus more interesting settings.

Our numerical results suggest that the number of maximally required layers for the compression scales quadratically in the number of bath sites for the considered system sizes. 
Since the number of two-qubit gates in each layer increases linearly with the number of bath sites $B$ and the total count of two-qubit gates for performing DMFT on a quantum computer scales with $B^3$.
This compression outperforms a Trotterized time evolution with the same gate count that is required for the training of the compressed circuits.

Based on our data, we thus find a quadratic dependence of the required number of layers for time evolution in the impurity model we consider. This is somewhat remarkable as, e.g. in \cite{Berthusen} the authors found an exponential dependency on the system size for the number of required layers for a time evolution in a Heisenberg chain. 
It is an open questions if this is caused by their chosen brick-wall ansatz or if the impurity system with a star geometry has a special feature so that the Hamiltonian variational ansatz is expressive enough with fewer layers.

Our results regarding the ground state preparation are also promising for the afore mentioned approaches of performing DMFT on a quantum computer that rely solely on measurements after the ground state approximation.
In these approaches, a large number of observables must be measured w. r. t. the ground state such that a high fidelity with the actual ground state is crucial.

Variational quantum algorithms are known to suffer from Barren plateaus. 
To circumvent this problem, we have chosen a Hamiltonian variational ansatz and a warm start regarding the parameters for the training. 
Nevertheless, the results in this paper have all been obtained using an error-free training, which is currently not feasible on current available hardware.
Therefore, in a real setup the error caused by the infidelities of the hardware and by the restricted number of shots would accumulate. The execution of our algorithm on real hardware would thus still require further improvement of achievable gate fidelities. 
Although we presented a method to reduce the circuit depth for training by introducing an additional energy measurement executed in parallel to the cost function measurement, further improvements are required to reduce the influence of hardware imperfections onto the training. 

As suggested in \cite{Zhao} it is possible to choose the step size for a Trotterization of the time evolution adaptively. 
This could be used to reduce the number of training steps for our algorithm and therefore to reduce the accumulated error but at the cost of more measurements between each cost function evaluation.
Nevertheless, as long as the training is classical feasible for small system sizes, our algorithm could be used to create circuits for DMFT test runs on hardware for solving the impurity Anderson model to study the influence of sampling and hardware noise on the results. These results could be used to improve the available and find new methods of classical post processing.

As stated in the introduction, classical methods like NRG and DMRG are currently the state-of-the-art method for solving impurity problems, and it is not expected that quantum computing approaches outperform these methods in the near term. However, since the computational cost of these solvers grows exponentially with the number of orbitals, cluster size, and entanglement, it is important to explore quantum algorithms whose scaling may ultimately be more favorable for multi-orbital and cluster DMFT problems.

\begin{acknowledgments}
This work was supported by the German Federal Ministry of Education and Research (BMBF) within the funding program “Quantum technologies — From basic research to market” in the projects MANIQU (Grant No. 13N15577) and EQUAHUMO (Grant No. 13N16067), and the Munich Quantum Valley, which is supported by the Bavarian state government with funds from the Hightech Agenda Bayern Plus. 
M.E.~is supported by the Cluster of Excellence ”CUI: Advanced Imaging of Matter“ of the Deutsche Forschungsgemeinschaft (DFG) – EXC 2056 – project ID 390715994.
\end{acknowledgments}


\bibliography{bib}

\addcontentsline{toc}{section}{References}

\clearpage
\onecolumngrid
\appendix
\pagenumbering{alph}

\begin{appendix}

\section{General Trotterization}
    \label{appendix:trotterization}
Here, the circuit for the Trotter step and therefore also the ansatz is discussed. We had the aim to let our circuit run on superconducting hardware therefore, we only considered single qubit rotations and two-qubit gates of adjacent qubits on quantum hardware. The Trotterization regarding $\hat{H}_\text{SIAM}$ after JWT is given by
\begin{align}
    \hat{U}_\text{imp}(\Delta t) &= e^{-i\Delta t \hat{H}_\text{imp}} = e^{-i\Delta t \frac{U}{4} (\mathds{1} + \hat{Z}_{0,\uparrow}\hat{Z}_{0,\downarrow})} \prod_\sigma e^{i\Delta t \left(\frac{\mu}{2}-\frac{U}{4} \right)\hat{Z}_{0,\sigma}}\\
    \hat{U}_\text{hyb}(\Delta t) &= e^{-i\Delta t \hat{H}_\text{hyb}} \approx \prod_{i=1,\sigma}e^{-i\Delta t\left(\frac{V_{i,\sigma}}{2} (\hat{X}_{0,\sigma}\hat{Z}_{1,\dots,i-1}\hat{X}_{i,\sigma} + \hat{Y}_{0,\sigma}\hat{Z}_{1,\dots,i-1}\hat{X}_{i,\sigma} \hat{Y}_{0,\sigma}\hat{X}_{i,\sigma})\right)}\\
    \hat{U}_\text{bath}(\Delta t) &= e^{-i\Delta t \hat{H}_\text{bath}} = \prod_{i=1,\sigma}e^{i\Delta t\frac{\epsilon_i}{2}\hat{Z}_{i,\sigma}}
\end{align}
The training is performed using fSim-Gates:
\begin{equation*}
    \text{fSim}(\theta,\phi)_{i,i+1} = \left(\begin{array}{cccc}
         1 & 0 & 0 & 0 \\
         0 & \cos(\theta) & -i\sin(\theta) & 0 \\
         0 & -i\sin(\theta) & \cos(\theta) & 0 \\
         0 & 0 & 0 & e^{-i\phi}
    \end{array} \right) 
\end{equation*}
For $\phi=0$ this leads to
\begin{equation*}
    \text{fSim}(\theta,0)_{i,i+1} \leftrightarrow  e^{-i\theta\left(\hat{X}_i\otimes\hat{X}_{i+1} + \hat{Y}_i\otimes\hat{Y}_{i+1}\right)} 
\end{equation*}
i. e., the terms appearing in $\hat{H}_\text{hyb}$ can be represented simply by these gates. However, because of the JWT, the exchanges with sites that are not adjacent to the impurity contain additional $\hat{Z}$ operators between. These exchange terms can be implemented by swapping the impurity with its direct neighbor using fermionic swap gates:
\begin{equation*}
    \text{fSwap}_{i,i+1} \leftrightarrow \left(\begin{array}{cccc}
        1 & 0 & 0 & 0  \\
        0 & 0 & 1 & 0 \\
        0 & 1 & 0 & 0 \\
        0 & 0 & 0 & - 1
    \end{array}\right)
\end{equation*}
These gates allow a swapping of qubit amplitudes while keeping track of the parity. This can be seen by expressing the fSwap-gate in terms of creation and annihilation operators:
\begin{equation*}
    \text{fSwap}_{i,j} = \mathds{1} + \hat{c}^\dag_i\hat{c}_j + \hat{c}^\dag_j\hat{c}_i - \hat{c}^\dag_i\hat{c}_i - \hat{c}^\dag_j\hat{c}_j
\end{equation*}
One can then show that the fSwap-gate fulfills the following relations
\begin{align*}
    \text{fSwap}_{i,j}\hat{c}^{(\dag)}_i\text{fSwap}_{i,j} &= \hat{c}^{(\dag)}_j\\
    \text{fSwap}_{i,j}\hat{c}^{(\dag)}_j\text{fSwap}_{i,j} &= \hat{c}^{(\dag)}_i\\
    \text{fSwap}_{i,j}\hat{c}^{(\dag)}_k\text{fSwap}_{i,j} &= \hat{c}^{(\dag)}_k \text{ with } k\neq i,j\\  
    \text{fSwap}_{i,j}\text{fSwap}_{i,j} = \mathds{1}
\end{align*}
The terms in $\hat{H}_\text{hyb}$ can then be transformed e. g. like
\begin{equation*}
    \hat{c}^\dag_i\hat{c}_{i+1} = \text{fSwap}_{i,i+1} \hat{c}^\dag_{i+1}\hat{c}_{i+2} \text{fswap}_{i, i+1}
\end{equation*}
so that after JWT each term can be transposed into two qubit gates. Furthermore, the number of two-qubit gates does not increase, since the fSwap can be incorporated into the fSim-gate in the following way:
\begin{equation*}
    \text{fSwap}_{i,i+1} \text{fSim}(\theta,0) = \text{fSim}_{i,i+1}(\theta+\frac{3}{2}\pi,0) \hat{Z}_i\otimes\hat{Z}_{i+1} \hat{S}_i \otimes \hat{S}_{i+1}=: R_\text{XX+YY}(\theta)
\end{equation*}
The terms of $\hat{H}_\text{imp}$ can also be expressed in terms of fSim-gates:
\begin{equation}
    \text{fSim}(0,\theta)_{i,j}\leftrightarrow e^{-i\theta (\mathds{1} + \hat{Z}_i\hat{Z}_j - \hat{Z}_i-\hat{Z}_j)}=:R_\text{ZZ}(\theta)
\end{equation}
The remaining parts of $\hat{H}_\text{imp}$ and $\hat{H}_\text{bath}$ can be expressed by using single qubit Z-rotations:
\begin{equation}
    \left(\begin{array}{cc} e^{-i\frac{\theta}{2}} & 0 \\ 0 & e^{i\frac{\theta}{2}}\end{array}\right) \leftrightarrow e^{-i\frac{\theta}{2}\hat{Z}}=: R_\text{Z}(\theta)
\end{equation}
A single Trotter step for $\hat{H}_\text{SIAM}$ in the second order decomposition chosen to be
\begin{equation}
    \hat{U}_\text{trotter}(\Delta t)= \hat{U}_\text{imp+bath}\left(\frac{\Delta t}{2}\right)\hat{U}_\text{hyb}(\Delta t)\hat{U}_\text{imp+bath}\left(\frac{\Delta t}{2}\right).
\end{equation}
As described in the main text, the required swapping of the impurity site along the qubit chain changes the representation of the lattice sites w. r. t. to the qubits. However, the Hadamard test, required for evaluating the Green's function and the cost function, requires an ancilla controlled application of a Pauli $X$ or $Y$ gate onto the impurity qubit and therefore that these two qubits are adjacent in the case of superconducting hardware. To circumvent this is issue we combine two Trotter steps in our definition \ref{trotterstepdefinition} and reveres the application order of $\hat{U}_\text{hyb}$. That way the qubit ordering is always brought back into its original form without additional overhead.
The strategy for the application of a Trotter step is, therefore, 
\begin{enumerate}
    \item start with the impurity next to each other and apply the impurity term and also the bath terms
    \item apply the hybridization terms combined with a fermionic swap of the two qubits
    \item repeat depending on the order of the Trotter Suzuki decomposition
\end{enumerate}
This is followed by a finial application of the impurity terms. This application scheme has proven to be the best for evaluating the terms in the Greens function. As one can see, the last gates in the Trotter step are followed by the starting and ending the time evolution gates with impurity term.

\section{Cost function}
    \label{appendix:costfunction}
We show that the global cost function
\begin{equation}
    C(\vec{\theta}) = 1 - \braket{\hat{W}^\dag(\vec{\theta}_n)|\hat{Z}_\text{anc}\otimes \left(\ket{0}\bra{0}
    \right)^{2B + 2}|\hat{W}(\vec{\theta}_n)}
    \label{appendixCostFunction}
\end{equation}
in fact return the correct circuits. 
The measure of $\hat{Z}_\text{anc}\otimes \left(\ket{0}\bra{0}\right)^{2B + 2}$ returns the expectation value
\begin{equation}
    \text{Re}\left(\braket{\hat{V}^\dag(\vec{\theta}_{n})\hat{U}(\Delta t) \hat{V}(\vec{\theta}_{n-1})}\braket{\hat{X}_{0,\sigma}\hat{V}^\dag(\vec{\theta}_{n-1})\hat{U}^\dag(\Delta t) \hat{V}(\vec{\theta}_{n})\hat{X}_{1\sigma}}\right)
    \label{appendixRealCost}
\end{equation}
where both expectations values are given w. r. t. the ground state prepared by $\hat{U}(\vec{\theta}_\text{gs})\ket{0} := \ket{GS} := \sum_i\alpha_i \ket{\alpha_i}$ with $\ket{\alpha_i}$ being states in the computational basis. 
Before training, the states after applying our training circuits to the ground state read
\begin{align}
    \hat{V}^\dag(\vec{\theta}_{n-1})\hat{U}^\dag(\Delta t) \hat{V}(\vec{\theta}_{n})\ket{\text{GS}} &= \sum_i\beta_i \ket{\alpha_i}  \\
    \hat{X}_{0\sigma}\hat{V}^\dag(\vec{\theta}_{n-1})\hat{U}^\dag(\Delta t) \hat{V}(\vec{\theta}_{n})\hat{X}_{0\sigma}\ket{\text{GS}} &=
     \sum_i\gamma_i\ket{\alpha_i} 
    + \sum_j \delta_j\ket{\delta_j}
\end{align}
where $\ket{\delta_i}$ are some states that are not included in the ground state, such that $\braket{\text{GS}|\delta_j} = 0$ for all $j$. 
This is only the case for the circuit including the $\hat{X}_{0\sigma}$ since the ansatz is excitations conserving. 
Multiplying both states with $\bra{\text{GS}}$ then returns $\beta:=\sum_i\alpha_i^*\beta_i$ and $\gamma:=\sum_i\alpha_i^*\gamma_i$ and our expression \ref{appendixRealCost} is equal to $\text{Re}(\beta^*\gamma)$. 
We can rewrite $\beta = |\beta|(\cos(\phi_1) + i\sin(\phi_1))$ and $\gamma = |\gamma|(\cos(\phi_2) + i\sin(\phi_2))$ for some angles $\phi_1$ and $\phi_2$ so that
\begin{align}
\begin{split}
    \text{Re}(\beta^*\gamma) &= |\beta||\gamma|\left(\cos(\phi_1) \cos(\phi_2) + \sin(\phi_1)\sin(\phi_2)\right) \\
    &= |\beta||\gamma| \cos(\phi_1 - \phi_2)
\end{split}
\end{align}
By enforcing the former expression to be equal to one (as it is done in the cost function)
\begin{align}
    |\beta||\gamma|\cos(\phi_1-\phi_2) &\overset{!}{=} 1 \\
    \cos(\phi_1-\phi_2) &\overset{!}{=} \frac{1}{|\beta||\gamma|}
\end{align}
Since $0<|\beta|\leq 1$ and $0<|\gamma|\leq 1$ it follows that $\frac{1}{|\beta||\gamma|}\geq1$. At the same time, $\cos(\phi_1-\phi_2)\in[-1,1]$ so that the condition can only be fulfilled, if 
\begin{align*}
    \frac{1}{|\beta||\gamma|} \overset{!}{=} 1 &\Rightarrow |\beta| = \frac{1}{|\gamma|} \Rightarrow |\beta| = 1 = |\gamma| \text{ and } \\
    \cos(\phi_1 - \phi_2) \overset{!}{=} 1 &\Rightarrow \phi_1 = \phi_2 + 2\pi n \text{ with } n \in \mathds{N}
\end{align*}
Thus, minimizing the cost functions leads to $\sum_i\alpha_i^*\beta_i = 1 = \sum_i\alpha_i^*\gamma_i$ such that the trained circuit $\hat{V}(\vec{\theta}_n)$ inverses the forward time evolution $\hat{U}(\Delta t) \hat{V}(\vec{\theta}_{n-1})$ w. r. t. to the states $\ket{\text{GS}}$ and $\hat{X}_{0\sigma}\ket{\text{GS}}$ without global phase difference between the two states. 
Therefore, the parametrized circuit can be used for the Green's function without further adjustments.

\section{Cost function reduction}
    \label{appendix:costfunctionreduction}
The Hamiltonian variational ansatz is a parametrized version of the gate sequence obtained from Trotterization (see \ref{trotterstepdefinition} in the main text). 
This ansatz is not only conserving in regard of the excitation number. 
For the local version of the cost function, it is required to measure the observable $\sum_{p,\sigma} \hat{Z}_{p,\sigma}$.
Here we show that this observable commute with the Hamiltonian of the single impurity Anderson model and thus with the Hamiltonian variational ansatz. 
It is trivial that $[\hat{H}_\text{imp},\sum_{p,\sigma}\hat{Z}_{p,\sigma}] = 0$ and $[\hat{H}_\text{bath},\sum_{p,\sigma}\hat{Z}_{p,\sigma}]= 0$.
Therefore, we are left with showing $[\hat{H}_\text{hyb},\sum_{p,\sigma}\hat{Z}_{p,\sigma}]=0$.
We consider the commutator for a single site:
\begin{align}
\begin{split}
    \left[\hat{X}_{0,\sigma'}\hat{Z}_{1,\sigma'}\dots\hat{Z}_{k-1,\sigma'}\hat{X}_{k,\sigma'},\sum_{p,\sigma}\hat{Z}_{p,\sigma}\right] &= \left[\hat{X}_{0,\sigma'},\hat{Z}_{0,\sigma'}\right]\hat{Z}_{1,\sigma'}\dots\hat{Z}_{k-1,\sigma'}\hat{X}_{k,\sigma'} + \hat{X}_{0,\sigma'}\hat{Z}_{1,\sigma'}\dots\hat{Z}_{k-1,\sigma'}\left[\hat{X}_{k,\sigma'},\hat{Z}_{k,\sigma'}\right] \\
    &= -2i \left(\hat{Y}_{0,\sigma'}\hat{Z}_{1,\sigma'}\dots\hat{Z}_{k-1,\sigma'}\hat{X}_{k,\sigma'} + \hat{X}_{0,\sigma'}\hat{Z}_{1,\sigma'}\dots\hat{Z}_{k-1,\sigma'}\hat{Y}_{k,\sigma'}\right)
    \label{costredXterm}
\end{split}
\\
\begin{split}
    \left[\hat{Y}_{0,\sigma'}\hat{Z}_{1,\sigma'}\dots\hat{Z}_{k-1,\sigma'}\hat{Y}_{k,\sigma'},\sum_{p,\sigma}\hat{Z}_{p,\sigma}\right] &= \left[\hat{Y}_{0,\sigma'},\hat{Z}_{0,\sigma'}\right]\hat{Z}_{1,\sigma'}\dots\hat{Z}_{k-1,\sigma'}\hat{Y}_{k,\sigma'} + \hat{Y}_{0,\sigma'}\hat{Z}_{1,\sigma'}\dots\hat{Z}_{k-1,\sigma'}\left[\hat{Y}_{k,\sigma'},\hat{Z}_{k,\sigma'}\right] \\
    &= 2i \left(\hat{X}_{0,\sigma'}\hat{Z}_{1,\sigma'}\dots\hat{Z}_{k-1,\sigma'}\hat{Y}_{k,\sigma'} + \hat{Y}_{0,\sigma'}\hat{Z}_{1,\sigma'}\dots\hat{Z}_{k-1,\sigma'}\hat{X}_{k,\sigma'}\right)
    \label{costredYterm}
\end{split}
\end{align}
Since \ref{costredXterm} and \ref{costredYterm} cancel each other for any $k$ it follows that $[\hat{H}_\text{hyb},\sum_{p,\sigma}\hat{Z}_{p,\sigma}]=0$.
Thus, 
\begin{equation}
\left[\hat{L}_n, \sum_{p,\sigma}\hat{Z}_{p,\sigma}\right]= \left[\hat{K}_n, \sum_{p,\sigma}\hat{Z}_{p,\sigma}\right]= 0
\end{equation}
but
\begin{equation}
    \left[\hat{U}_\text{GS}, \sum_{p,\sigma}\hat{Z}_{p,\sigma}\right] \neq 0.
\end{equation}
While the observable commutes with the parametrized circuit, it does not with initial layer of gates for the introduction of the excitations into the system. 
This is crucial for the local cost \ref{localCostFunction}.
Consider $\hat{U}_\text{GS} = \hat{U}'_\text{GS}\hat{U}_\text{ini}$ where $\hat{U}'_\text{GS}$ is the Hamiltonian variational ansatz and $\hat{U}_\text{ini}$ the gate sequence required for the correct excitation number. 
In the following, we show that it is not possible to find a $\hat{U}_\text{ini}\neq \mathbf{1}$ that introduces the correct number of excitations and fulfills $\left[\hat{U}_\text{ini}, \sum_{p,\sigma}\hat{Z}_{p,\sigma}\right]=0$.
Consider $\hat{U}_\text{ini}$ with
\begin{equation*}
    \hat{U}_\text{ini}\ket{\vec{0}} = \ket{n_\uparrow, n_\downarrow}
\end{equation*}
where $n_\sigma$ are the number of excitations for spin $\sigma$ and $n_\uparrow + n_\downarrow \geq 1$. 
Following the application of $\hat{U}_\text{ini}$ with $\sum_{p,\sigma}\hat{Z}_{p,\sigma}$ we have
\begin{equation}
    \sum_{p,\sigma}\hat{Z}_{p,\sigma}\hat{U}_\text{ini}\ket{\vec{0}} = 2m\ket{n_\uparrow, n_\downarrow} \text { with } m \in \{-(B+1), -B, \dots, -1, 0, 1, \dots B\} 
\end{equation}
Since $\ket{n_\uparrow,n_\downarrow}\neq\ket{\vec{0}}$ it follows that $m\neq B+1$.
However, 
\begin{equation}
    \hat{U}_\text{ini}\sum_{p,\sigma}\hat{Z}_{p,\sigma}\ket{\vec{0}} = 2(B+1)\hat{U}_\text{ini}\ket{\vec{0}} = 2(B+1)\ket{n_\uparrow,n_\downarrow} \neq 2m\ket{n_\uparrow,n_\downarrow} \text{ for any } m \in \{-(B+1), -B, \dots, -1, 0, 1, \dots B\}.
\end{equation}
Thus, $\left[\hat{U}_\text{ini}, \sum_{p,\sigma}\hat{Z}_{p,\sigma}\right] \neq 0$.
The local cost function \ref{localCostFunction} contains the term 
\begin{equation}
\hat{U}_\text{GS}\sum_{p,\sigma}\hat{Z}_{p,\sigma} \hat{U}^\dag_\text{GS} = \hat{U}'_\text{gs}\hat{U}_\text{ini}\sum_{p,\sigma}\hat{Z}_{p,\sigma}\hat{U}_\text{ini}^\dag\hat{U}'^\dag_\text{GS}.
\end{equation}
We now consider two cases. 
First, $n_\uparrow = n_\downarrow$. In that case $\hat{U}_\text{ini}$ consists of $\hat{X}$-gates applied to a subset of the qubits $\mathcal{M}_\text{ini}$, i. e. $\hat{U}_\text{ini} = \prod_{j\in\mathcal{M}_\text{ini}} \hat{X}_j$.
We now use the fact that $\hat{X}_j\hat{Z}_j\hat{X}_j = 2\hat{X}_j\hat{Z}_j\hat{X}_j - \hat{X}_j\hat{Z}_j\hat{X}_j = 2\hat{X}_j\hat{Z}_j\hat{X}_j + \hat{Z}_j$ thus
\begin{equation}
    \hat{U}_\text{ini}\sum_{p,\sigma}\hat{Z}_{p,\sigma}\hat{U}_\text{ini}^\dag = \sum_{p,\sigma}\hat{Z}_{p,\sigma}\hat{U}_\text{ini}\hat{U}_\text{ini}^\dag + 2\hat{U}_\text{ini}\sum_{j\in\mathcal{M}_\text{ini}}\hat{Z}_j\hat{U}_\text{ini}^\dag = \sum_{p,\sigma}\hat{Z}_{p,\sigma} + 2\hat{U}_\text{ini}\sum_{j\in\mathcal{M}_\text{ini}}\hat{Z}_j\hat{U}_\text{ini}^\dag
\end{equation}
In the second case, $n_\uparrow \neq n_\downarrow$, the initialization requires at least one gate sequence for two qubits $i$ and $j$ such that $\ket{0,0} \mapsto \frac{1}{\sqrt{2}}(\ket{1,0}+\ket{0,1})$.
This can be accomplished by the gate sequence $\hat{S}_j\text{fSim}\left(\frac{\pi}{4},0\right)_{i,j}\hat{X}_{i}$. 
Except for the initial $\hat{X}_{i}$ gate, the observable $\sum_{p,\sigma}\hat{Z}_{p,\sigma}$ commutes with this sequence, such that it can be traced back to the first case. 
The term in \ref{localCostFunction} therefore reduces to
\begin{equation}
\hat{U}_\text{GS}\sum_{p,\sigma}\hat{Z}_{p,\sigma} \hat{U}^\dag_\text{GS} = \sum_{p,\sigma} \hat{Z}_{p,\sigma}+ 2 \hat{U}_\text{GS}\sum_{j\in\mathcal{M}_\text{ini}}\hat{Z}_j\hat{U}^\dag_\text{GS}.
\end{equation}
Thus, the cost function can be rewritten as shown in \ref{finalCostAnc} - \ref{finalCostUgsZAnc}.

\section{Ansatz}
    \label{appendix:ansatz}
The design of the ansatz must fulfill certain requirements. 
First, it must be expressive enough to approximate the time evolution correctly. Furthermore, a excitations number conserving ansatz proved to be advantageous. 
As seen in equations \ref{finalCostAnc} - \ref{finalCostUgsZAnc} the training relays on the fact that the state after applying the training circuit has a large overlap with the desired state and that the amplitude of all states with a wrong number of excitations is minimized. 
Using an excitations number preserving ansatz, these amplitudes should be already low from the get go, so that the optimization step can be performed faster. 
A natural choice for the ansatz is therefore a Hamiltonian variational ansatz, that uses the same gate structure as the Trotter step. 
That way, the iterative training is compressing the Trotterized time evolution in fewer Trotter steps by adjusting the parameters of the gates accordingly. 
Another important feature for the ansatz is, that the final gate does not commute with the Pauli X- and Y-gate applied to the impurity site for measuring the Greens function. 
Consider e. g. the term
\begin{equation}
    \text{Re}(\braket{\hat{U}^\dag(\theta)\hat{X}_{i,\sigma}\hat{U}(\theta)\hat{X}_{i,\sigma}})
\end{equation}
If the ansatz ends with gates that commute with $\hat{X}_{i, \sigma}$ it would cancel with the first gates in $\hat{U}^\dag(\theta)$ not contributing to the expectation value.
However, because of the compression, each gate contributes more to the time evolution than it would in Trotter time evolution, so its cancellation would lead to the wrong value or phase.
This issue becomes more relevant the further in time the calculation is performed.
This is the reason why we choose to start and end the Trotter decomposition with the impurity term that does not commute with $\hat{X}_{i, \sigma}$.
We further removed some single qubit $R_z$-gates, as seen in figure \ref{ansatz} since they did not improve the performance. 
This way, the computational cost can be reduced.

\section{Parameter Shift Rule}
    \label{appendix:parametershiftrule}
The parameter shift rule has been introduced in \cite{Mitarai, Schuld} for single qubit Pauli rotation gates and generalized in \cite{Anselmetti} for multi qubit gates. Based on these results, this section shows that the parameter shift rule can also be applied to the cost function (32) used for the algorithm. It can be rewritten using the following definitions:
\begin{align*}
    V(\vec{\theta}_{n+1}) &:= \hat{U}^\dag(\vec{\theta}_n)\hat{U}^\dag(\Delta t)\hat{U}(\vec{\theta}_{n+1}) \\
    \hat{O} &:= \ket{\text{GS}}\bra{\text{GS}}\hat{X}_{i\sigma}\\
    \ket{\text{GS}_x} &:= \hat{X}_{i,\sigma}\ket{\text{GS}}
\end{align*}
The cost function is then given by
\begin{equation}
    C(\vec{\theta}_{n+1}) = 1- \text{Re}(\braket{GS|\hat{V}^\dag(\vec{\theta}_{n+1})\hat{O}\hat{V}(\vec{\theta}_{n+1})|\text{GS}_x}).
\end{equation}
and its derivative, by using the linearity of the derivative and the product rule, by
\begin{align}
    \frac{\partial }{\partial \theta^i}C(\vec{\theta}_{n+1}) &= - \left(\text{Re}(\braket{GS|\left(\frac{\partial}{\partial \theta^i}\hat{V}^\dag(\vec{\theta}_{n+1})\right)\hat{O}\hat{V}(\vec{\theta}_{n+1})|\text{GS}_x}) + \text{Re}(\braket{GS|\hat{V}^\dag(\vec{\theta}_{n+1})\hat{O}\left(\frac{\partial}{\partial \theta^i}\hat{V}(\vec{\theta}_{n+1})\right)|\text{GS}_x})\right)
\end{align}
The operator $\hat{V}$ consists of a product of unitary operators, each depending on a single parameter. Assuming $M$ different parameters, the derivative of $\hat{V}$ only involves one of these gates, such that
\begin{equation*}
    \frac{\partial}{\partial \theta^i}\hat{V}(\vec{\theta}_{n+1}) = \hat{U}^\dag(\vec{\theta}_{n})\hat{U}^\dag(\Delta t) \prod_{j = 1}^{i -1} \hat{U}(\theta^j)\frac{\partial}{\partial \theta^i}\hat{U}(\theta^i) \prod_{k = i + 1}^{M}\hat{U}(\theta^k)
\end{equation*}
The unitary operators which are applied before and after $\hat{U}(\theta^i)$ can be included into the definition $\hat{O}$, $\ket{\text{GS}_x}$ and $\ket{\text{GS}}$ to further simplify the expression:
\begin{equation}
    \frac{\partial}{\partial \theta^i}\hat{V}(\vec{\theta}_{n+1}) = - \left(\text{Re}(\braket{\text{GS}|\left(\frac{\partial}{\partial \theta^i}\hat{U}^\dag(\theta^i)\right)\hat{O}\hat{U}^\dag(\theta^i)|\text{GS}_x}) + \text{Re}(\braket{\text{GS}|\hat{U}^\dag(\theta^i)\hat{O}\left(\frac{\partial}{\partial \theta^i}\hat{U}^\dag(\theta^i)\right)|\text{GS}_x})\right)
\end{equation}
Now it can easily been seen that the parameter shift rule can be applied here. In the ansatz only gates of the type 
\begin{equation}
    \hat{U}(\theta^i) = \exp(-i\frac{\theta^i}{2}P)
\end{equation}
are used, where $P$ is the generator with $P \in \{\hat{Z}, \hat{Z}\otimes\hat{Z}, \hat{X}\otimes\hat{X} + \hat{Y}\otimes\hat{Y}\}$. These generators have either the eigenvalues $\{-1,1\}$ or $\{-1,0,1\}$ allowing the calculation of each partial derivative by evaluating two or four expectation values.

\section{Optimization Process}
\renewcommand{\thefigure}{\thesection.\arabic{figure}}
\setcounter{figure}{0}
\label{appendix:qcresources}
Since both the VQE for the preparation of the ground state and the time compression algorithm must be performed for each new iteration in the DMFT-loop, the efficiency of the optimization process is critical. Specifically, we examine the number of gradient evaluations required to minimize their respective cost functions.
Figures \ref{costfunctionevolutionGS} and \ref{costfunctionevolutiontime} (the latter averaged over all time steps) illustrate this process for the setup described and used in section \ref{subsec:dmftloopconvergence}.
The results presented are based on state-vector simulations comparing two optimization algorithms: BFGS and ADAM.
BFGS converges faster by approximating the Hessian matrix through previous gradient measurements. However, it is generally ill-suited for the noisy gradients inherent in hardware executions.
ADAM is a robust gradient-descent-based optimizer. Here, we utilize a learning rate of $\eta = 0.1$. While ADAM requires more gradient evaluations than BFGS, it offers the advantage of not needing to evaluate the cost function itself. 
To reach a target fidelity of $10^{-4}$, the ground state preparation requires approximately 35 and 100 gradient evaluations using BFGS and ADAM, respectively. 
For the time evolution compression, these numbers decrease in both cases, to 14 for the ground state and 35 for the time evolution.
In practical hardware setting, one would typically begin with "rough" gradient estimations using a low shot count, gradually increasing the precision as convergence nears. 
Furthermore, the total number of evaluations could likely be reduced in subsequent DMFT iterations, by using the optimized parameters from the previous step as a starting point.
It should be noted that because values obtained via the parameter-shift rule are generally larger than the cost function values themselves, the final gradient evaluations typically require between $10^5$ and $10^6$ shots to maintain accuaracy.
\begin{figure}[h]
    \centering
    \begin{minipage}{0.49\textwidth}
    \includegraphics[width=1\textwidth]{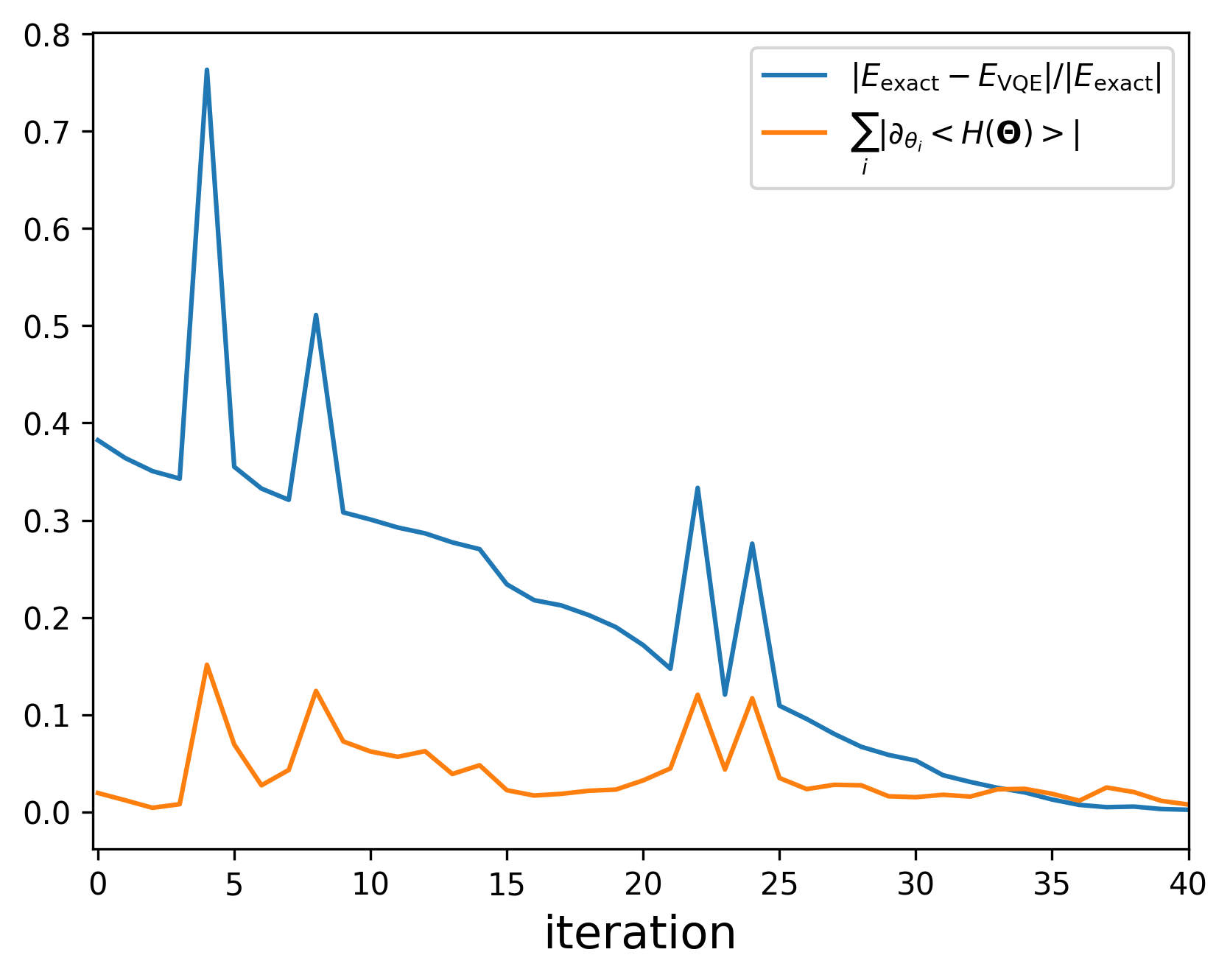}
    \end{minipage}
    \begin{minipage}{0.49\textwidth}
        \includegraphics[width =1\textwidth]{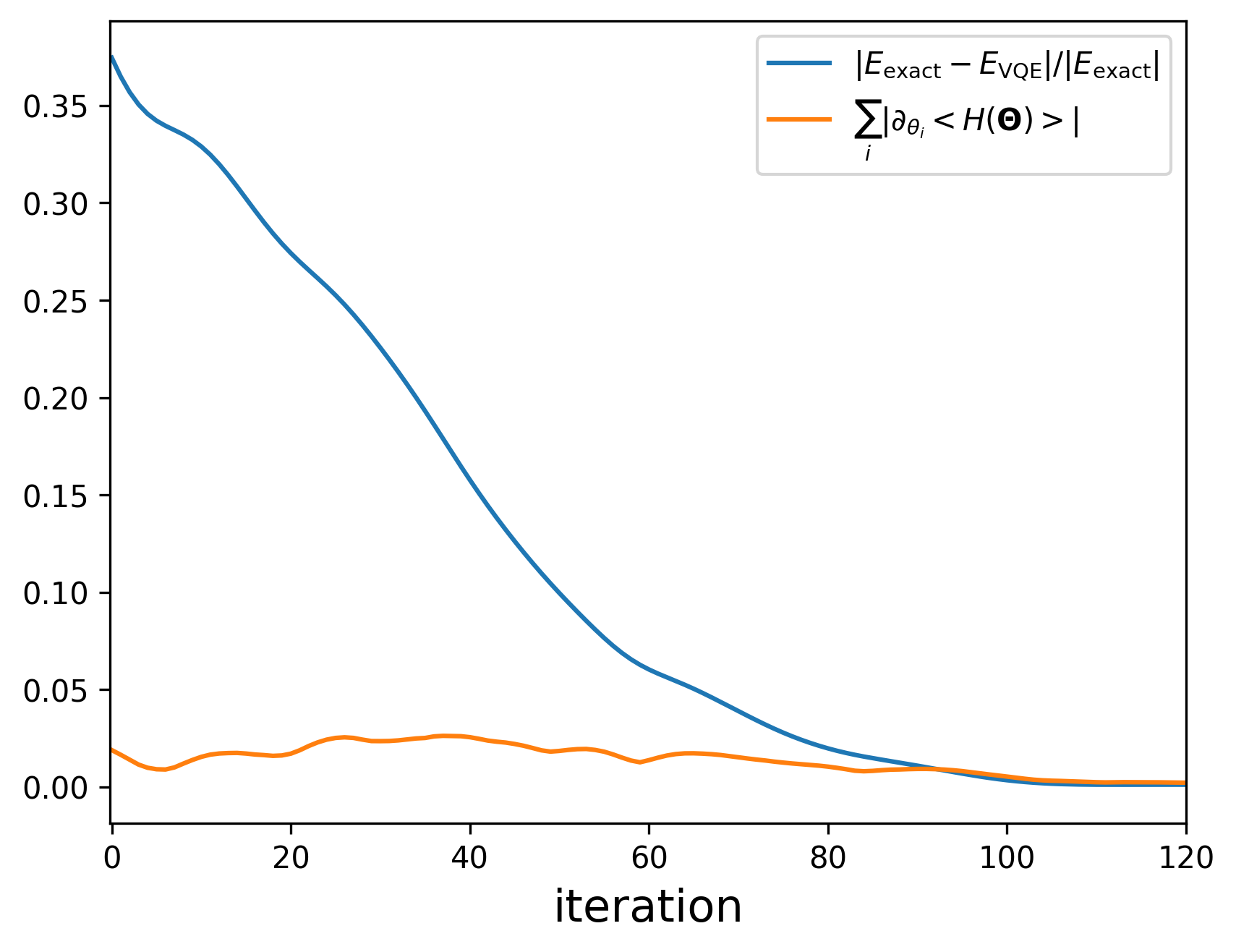}
    \end{minipage}
    \caption{Evolution of the ground state energy and the gradient for an ansatz with $L=2$ during the optimization process with BFGS (left) and ADAM (right) for a system with $U=4v$, $B=2$ and $n=0.5$.}
    \label{costfunctionevolutionGS}
\end{figure}

\begin{figure}[h]
    \centering
    \begin{minipage}{0.49\textwidth}
    \includegraphics[width=1\textwidth]{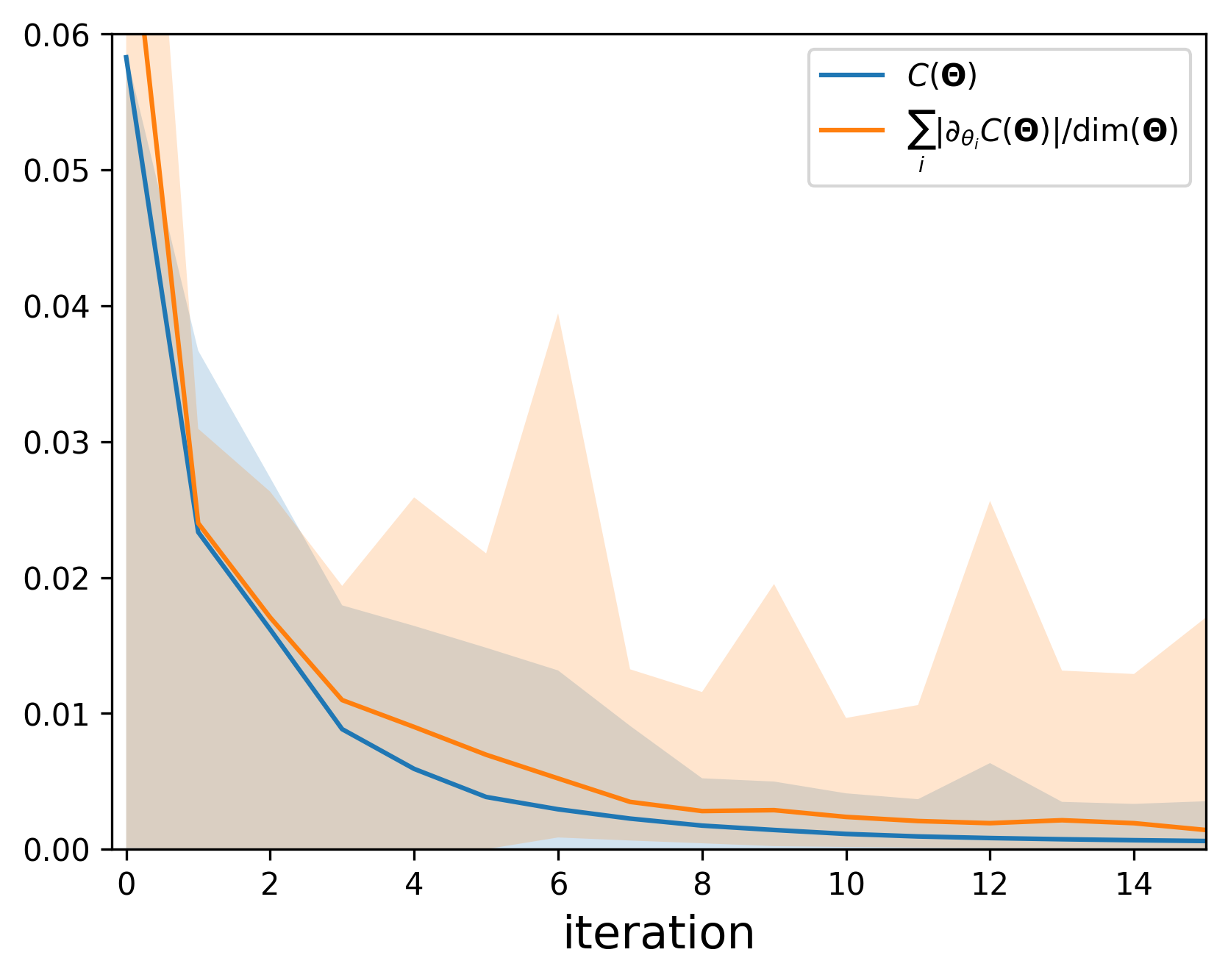}
    \end{minipage}
    \begin{minipage}{0.49\textwidth}
        \includegraphics[width =1\textwidth]{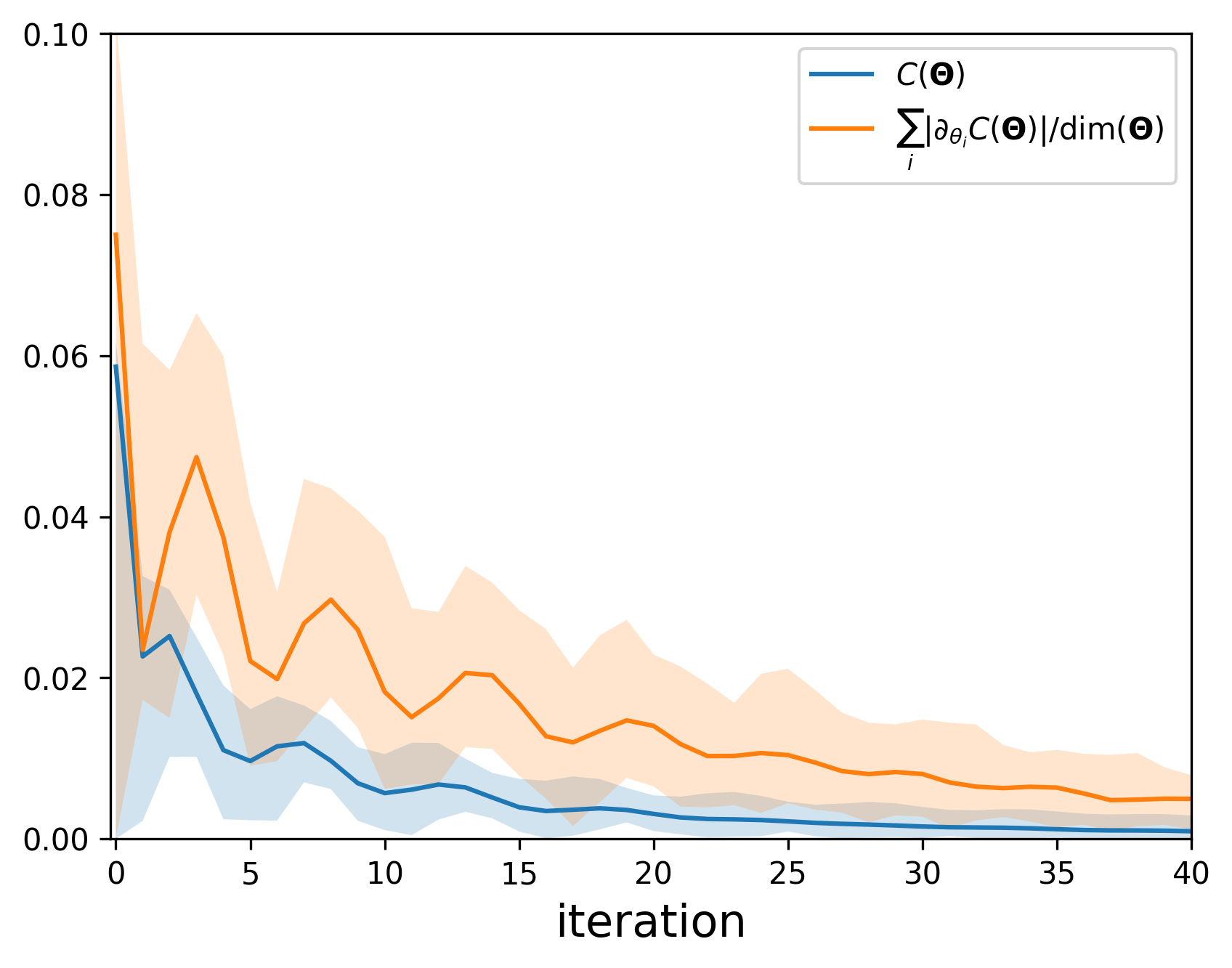}
    \end{minipage}
    \caption{Evolution of the cost function of the time compression algorithm and its gradient for an ansatz with $L=3$ during the optimization process with BFGS (left) and ADAM (right) for a system with $U=4v$, $B=2$ and $n=0.5$. We illustrate the average over all 500 simulated time steps where the shaded area illustrates the range of the appearing values.}
    \label{costfunctionevolutiontime}
\end{figure}

\section{Fitting procedure of the Greens Function}
    \label{appendix:greensfit}
Normally, after calculating the Green's function $G^R_{\text{imp}, \sigma}(t)$ the Fourier transformation is performed to obtain $G^R_{\text{imp}, \sigma}(\omega)$:
\begin{equation}
    G^R_{\text{imp}, \sigma}(\omega) = \int_{-\infty}^\infty dt e^{i\omega t} G^R_{\text{imp}, \sigma}(t)
\end{equation}
This Fourier transformation is only well defined if the time dependent Green's function decays in time for $t \rightarrow\pm \infty$. Although the retarded Green's function is zero for $t < 0$ by definition, it does not decay for $t \rightarrow \infty$ in general. Therefore, the real frequency $\omega$ must be replaced by a complex one $\omega + i\eta$ where $\eta > 0$ is an infinitesimal small number, such that
\begin{equation}
    G^R_{\text{imp}, \sigma}(\omega+i\eta) = \int_{-\infty}^\infty dt e^{i\omega t} e^{-\eta t} G^R_{\text{imp}, \sigma}(t)
\end{equation}
Nevertheless, performing this Fourier transformation is only possible if $G^R_{\text{imp}, \sigma}(t)$ is calculated far enough in time. However, this is not possible with the current state of quantum computing. Instead we rewrite the Green's function in terms of the Lehmann parameter:
\begin{equation}
    G^R_{\text{imp},\sigma}(t) = \sum_j |\braket{j|\hat{c}^\dag_{i,\sigma}|\text{GS}}|^2e^{-i(E_j - E_\text{GS})t} + |\braket{j|\hat{c}_{i,\sigma}|\text{GS}}|^2 e^{i(E_j-E_\text{GS})t}
\end{equation}
where $\ket{j}$ are the eigenstates of $\hat{H}_\text{SIAM}$ with eigenenergy $E_j$. After introducing the Lehmann parameter,
\begin{align}
    \alpha_j &= |\braket{j|\hat{c}^\dag_{i,\sigma}|\text{GS}}|^2\\
    \beta_j &= |\braket{j|\hat{c}_{i,\sigma}|\text{GS}}|^2 \\
    \omega_j &= E_j - E_\text{GS}
\end{align}
we split the Green's function into its real and imaginary components:
\begin{align}
    G^R_{\text{imp},\sigma}(t) &= \sum_j \alpha_je^{-i\omega_jt} + \beta_je^{i\omega_jt} \\
    &= \sum_j(\alpha_j + \beta_j)\cos(\omega_jt) + i(\beta_j - \alpha_j)\sin(\omega_jt)\\
    &= \sum_j\gamma_j\cos(\omega_jt) + i \delta_j\sin(\omega_jt)
\end{align}
with $\gamma_j := \alpha_j + \beta_j$ and $\delta_j := \beta_j - \alpha_j$. Now, the parameters $\gamma_j, \delta_j, \omega_j$ can be fitted onto the real and imaginary component of the Green's function, respectively, and then used to obtain the Lehmann parameters with
\begin{align*}
    \alpha_j &= \frac{1}{2}(\gamma_j - \delta_j)\\
    \beta_j &= \frac{1}{2}(\gamma_j + \delta_j)
\end{align*}
However, since only a small number of points can be evaluated, the fitting procedure is not precise enough. Therefore, further conditions are applied on the Lehmann parameters (see \cite{Rungger}):
\begin{align}
    \sum_j \alpha_j + \beta_j &\overset{!}{=} 1\\
    \sum_j \frac{\alpha_j}{\epsilon_i - \omega_j} + \frac{\beta_j}{\epsilon_i + \omega_j} &\overset{!}{=} 0 \quad \forall i \in \{1,\dots,B\} \\
    \sum_j \frac{\alpha_j}{(\epsilon_i - \omega_j)^2} + \frac{\beta_j}{(\epsilon_i + \omega_j)^2} &\overset{!}{=} \frac{1}{V_i^2} \quad \forall i \in \{1,\dots,B\}
\end{align}
These are included as additional terms in the cost function for the fitting procedure to enforce fitting parameters that are closer to the actual Lehmann parameters.

\end{appendix}

\end{document}